\documentclass[authoryear,12pt]{elsarticle}

\usepackage{lineno,hyperref}
\usepackage{amsmath}
\modulolinenumbers[5]
\usepackage{placeins}
\usepackage[top=0.7in,bottom=0.7in,left=0.7 in, right = 0.7 in]{geometry} 

\makeatletter
\def\ps@pprintTitle{%
 \let\@oddhead\@empty
 \let\@evenhead\@empty
 \def\@oddfoot{\centerline{\it{Preprint, 30 September 2021}}}%
 \let\@evenfoot\@oddfoot}
\makeatother

\biboptions{authoryear}

\usepackage{subfig} 
\usepackage{epstopdf} 
\usepackage{amsmath,amssymb,amsthm,amsfonts} 
\usepackage{varioref}
\usepackage{doi}
\usepackage{booktabs}
\usepackage{caption}
\usepackage{multirow}
\usepackage{longtable}
\usepackage{pdflscape}
\usepackage{xcolor}
\usepackage{rotating}
\usepackage{lscape}

\usepackage{setspace} 

\usepackage{verbatim}

\usepackage{adjustbox}

\usepackage{booktabs}

\theoremstyle{definition}

\theoremstyle{plain}

\theoremstyle{plain}

\theoremstyle{plain}
\newtheorem{remark}{Remark}

\theoremstyle{plain}

\begin{document}
	\onehalfspacing
		
\title{Cyber Risk Frequency, Severity and Insurance Viability}


  \author[add1]{Matteo Malavasi}
  \ead{Matteo.Malavasi@mq.edu.au}
    \author[add2,add1]{ Gareth W. Peters}
  \ead{garethpeters@ucsb.edu}   
  \author[add1]{Pavel V. Shevchenko\corref{cor1}}
  \ead{Pavel.Shevchenko@mq.edu.au}  
  \author[add1]{Stefan Tr\"uck}
  \ead{Stefan.Trueck@mq.edu.au}  
  \author[add1]{Jiwook Jang}
  \ead{Jiwook.Jang@mq.edu.au}
  \author[add3]{Georgy Sofronov}
  \ead{Georgy.Sofronov@mq.edu.au}

  \cortext[cor1]{Please address correspondence to Pavel Shevchenko}
  \address[add1]{Department of Actuarial Studies and Business Analytics, Macquarie University, Australia}
  \address[add2]{Department of Statistics and Applied Probability, University of California Santa Barbara, USA}
  \address[add3]{Department of Mathematics and Statistics, Macquarie University, Australia}

\begin{abstract}
In this study an exploration of insurance risk transfer is undertaken for the cyber insurance industry in the United States of America, based on the leading industry dataset of cyber events provided by Advisen. We seek to address two core unresolved questions. First, what factors are the most significant covariates that may explain the frequency and severity of cyber loss events and are they heterogeneous over cyber risk categories? Second, is cyber risk insurable in regards to the required premiums, risk pool sizes and how would this decision vary with the insured companies industry sector and size?
We address these questions through a combination of regression models based on the class of Generalised Additive Models for Location Shape and Scale (GAMLSS) and a class of ordinal regressions. These models will then form the basis for our analysis of frequency and severity of cyber risk loss processes. We investigate the viability of insurance for cyber risk using a utility modelling framework with premium calculated by classical certainty equivalence analysis utilising the developed regression models. Our results provide several new key insights into the nature of insurability of cyber risk and rigorously address the two insurance questions posed in a real data driven case study analysis.

{\it\noindent Keywords: cyber risk, GAMLSS, cyber risk insurance, ordinal regression}
\end{abstract}

\maketitle

\newpage
\section{Introduction}
The understanding, mitigation, reporting, risk management and insurance modelling related to cyber risk are still in the early stages of development. However, with the increasing focus on Information Technology (IT) and cyber related risk that many companies, governments and regulators are beginning to actively explore in all industry sectors, these aspects of cyber risk are coming to the forefront of the insurance industry and risk managers portfolios. In this work we seek to study the areas that are under explored in regards to insurance and loss modelling aspects of cyber risk.

In modern business practices the increasing intersection between automation and efficiency and the enhanced uptake of IT infrastructure, occurring in all areas of business, industry and service sector processes has resulted in a greater impact of cyber risk. This has driven an enhanced focus on cyber security and motivated a much stronger cyber risk awareness in both practice, governance, regulation and system design. This will continue to naturally progress due to the fact that organisations of all sizes in both the public and private sectors are increasingly reliant on IT systems in order to execute business processes that support the delivery of services. If there is a breakdown or failure in these systems, the organisation will experience a direct negative impact on the processes it supports, resulting in reduction of service and disruptions that ultimately impact on the organisations ability to meet its objectives. This in turn can generate a variety of losses that are both directly and indirectly attributable to cyber events. In \cite{eling2016we,peters2018understanding,eling2020cyber} and book length discussions \cite{refsdal2015cyber,bouveret2018cyber} several detailed perspectives of the current state of the cyber risk insurance market are presented. In this study we seek to extend the focus by utilising a quantitative model based framework to assess critical questions regarding cyber risk modelling and risk transfer efficiency, effectiveness and availability. Such statistical modelling perspectives require a reliable data source, which is a challenge to obtain in the cyber risk context. We utilise the leading industry dataset from Advisen\footnote{\url{https://www.advisenltd.com/data/cyber-loss-data/}} (hereafter referred to as Advisen Cyber Loss Data) that maintains a detailed account of major cyber risk losses across all industry sectors around the world. Our analysis is focused on the United States of America (USA) experience as this represents the most comprehensive and complete set of records in this dataset and we believe the findings we obtain naturally transfer to other developed economies.

In this work we seek to make novel contributions to the modelling and quantification of cyber risk along two key lines of questioning. First, what factors are the most significant explanatory variables that may explain the frequency or severity of cyber loss events and are these factors heterogeneous over cyber risk categories? Second, is cyber risk insurable in regards to required premiums, risk pool sizes, and how would this decision vary with the insured company industry sector and size?

Regarding the modelling of cyber risk, there are studies that investigate basic statistical properties of cyber loss data in a non-regression framework, such as \cite{biener2015insurability}, \cite{eling2015modelling}, \cite{eling2017data}, \cite{eling2018copula} and \cite{peters2018statistical}. However, none of these studies has applied such a wide range of flexible regression models to an extensive dataset of losses from cyber events as we do. Therefore, our analysis will be one of the first to allow the thorough examination of key risk drivers of losses from cyber risk for different industry sectors and event types. 
In regards to questions of insurability several authors have addressed aspects of the insurability of cyber risk losses, the efficiency and structuring of insurance for such risk types; see, \cite{peters2011impact}, \cite{mukhopadhyay2013cyber}, \cite{biener2015insurability}, \cite{camillo2017cyber}, \cite{eling2019actual}, and \cite{romanosky2019content}. In this work we seek to explore insurability from a data driven quantitative perspective rather than an economic risk transfer perspective which has been undertaken in the aforementioned works. By this we aim to provide new perspectives and insights on the question of insurability of cyber risk. 

Our analysis is uniquely based on a combination of carefully developed regression models that can adequately accommodate factors and key risk drivers that may influence loss frequency and loss severity in cyber risk loss modelling through generalisations of the well established Generalised Linear Modelling (GLM) regression framework to the Generalised Additive Models for Location Shape and Scale (GAMLSS) pioneered by \cite{stasinopoulos2007generalized}. This class of models allows one to identify and associate the risk drivers or regression factors that influence the frequency and severity of cyber loss processes in the mean, variance, and unlike standard GLM regression models, we can also associate regression factors to higher order moments such as skewness and kurtosis. Effectively, the use of GAMLSS regression for modelling the severity and frequency of losses in cyber risk can prove to be a powerful framework in better understanding which factors are influential in the central moments of studied loss processes. Importantly, we can distinctly identify which factors are influential in the tail behaviour of cyber losses. This is an advantage over GLM regression modelling that is particularly poignant when heavy tailed loss processes are under consideration. Then to ensure we have meaningful results not directly influenced by leverage effects arising from the sheer magnitude of the cyber losses studied in some risk lines, we also undertake an analysis of the dataset using an ordinal rank based regression, using a framework proposed in \cite{giudici2020cyber}. This framework allows us to develop meaningful insights into concordance rankings for risk profiles in cyber risk loss categories not attainable readily from the standard GLM regression structures.

We conclude our study by focusing on the selected optimal regression structures to undertake an analysis of the insurability of cyber risk losses based on the developed frequency and severity regression models. We study the premium and risk pool size for insurance mitigation, using a utility based framework with certainty equivalence analysis for risk averse firms. The manner in which we have performed the analysis also allows us to provide insights with respect to the risk appetite and insurance cost on a firm by firm basis. This enhances the types of economic analysis undertaken in work such as \cite{lis2019cyberattacks}.

The paper is structured as follows. The model framework is defined in Section~\ref{sec:model}, while our dataset is described in Section~\ref{sec:data}. Sections ~\ref{Sec_gamlss_analysis} and \ref{Sec:ranked_based_regression} are devoted to GAMLSS and rank based regressions respectively.  Applications for Value-at-Risk and insurance premium calculations are presented in Section \ref{Sec:case_study}. The paper is concluded in Section~\ref{Sec:conclusions}. Additional statistical results regarding various model fits are presented in the Appendix. 

\section{Model Framework}
\label{sec:model}
Cyber risk related monetary losses (hereafter losses), as well as more general IT related Operational Risk (OpRisk) losses, exhibit extreme events. It is well documented in the literature that in the presence of a heavy tailed severity distribution, standard ordinary least squares (OLS) based techniques might not be appropriate \cite[see, among others,][]{cope2008,dahen2010,ganegoda2013,chavez2016}. Moreover, cyber risk affects a variety of different actors, in many different ways such that flexible modelling approaches are needed in order to draw meaningful inference on the driving risk factors \cite[see, e.g.][] {peters2018statistical}. 

A common statistical modelling framework used widely in OpRisk is the Loss Distribution Approach (LDA), in which risk related losses are a marked point process, occurring at random points in time. One can then model the frequency component with a counting process and the severity or size of the losses incurred over time at the times of events (the marks) with a positively supported heavy tailed loss distribution model; see e.g. \cite{shevchenko2011modelling,cruz2015fundamental}. One popular approach adopted in cyber risk modelling has been to utilise a Generalised Pareto model that is obtained from the framework of Extreme Value Theory (EVT) via a classical estimation approach of the peaks-over-threshold (POT) method. This will accommodate the tail behaviour to adequately explain and allow for the modelling of extreme events encountered in cyber risk loss modelling \cite[see, among others,][]{ganegoda2013,chavez2016,eling2019actual}. 

To setup an LDA modelling framework with frequency and severity components, we establish the following notation. Let $N$ be a discrete counting random variable for the number of loss events in a specified period of time (typically annual) and let $\widetilde{Y}_i$, $i=1,\dots,N$, be  positive loss random variables, defined on a common probability space $\left(\Omega,\mathcal{F},\mathbb{P}\right)$, obtained from a severity loss distribution. Then the aggregated loss or total loss over the specified period will be denoted by the compound random variable $Z$ defined as:
\begin{equation*}
    Z = \sum_{i=1}^N\widetilde{Y}_i.
\end{equation*}
Typically, it is assumed that $N$ and $\widetilde{Y}_i$, $i=1,\dots,N$ are all independent. As outlined in this set up, the LDA approach aims to find frequency and severity distributions supported by the data, in order to provide insights on the risk under investigation. In practice, one may consider several choices for the frequency and severity models. Popular examples include modelling the event frequency $N$ by a Poisson distribution. If there is over- or under-dispersion in the observed frequency one may opt for a Negative Binomial or Binomial distribution, respectively; see discussion in the cyber risk modelling context in \cite{edwards2016, eling2019actual}. In terms of the severity or loss size model, in order to accommodate extreme events, the severity distribution can be modelled with a positively supported distribution that can accommodate heavy tailed features. 

\begin{remark} Due to the lack of good datasets for cyber risk losses, some authors suggest to avoid the use of classical statistical methods for estimation of the LDA frequency and severity distributions using historical data (especially for low-frequency/high-impact risks) and rather to rely on scenario analysis approaches. For example, \cite{rakes2012security} argue that for sparse/high-impact IT security breaches one can rely on an expert’s judgement defining worst-case scenarios and their likelihood. It is also clear that estimation based on historical losses is backward looking and it is challenging to account for a constantly changing environment. It has been a common practice in OpRisk modelling (in the banking industry) to use scenario analysis for estimation of frequency and severity distributions. Moreover, there was a Basel II regulatory requirement for bank's internal OpRisk capital models to include  the use of internal data, external data, scenario analysis and factors reflecting the business environmental and internal control systems; see \cite{BCBS2006}. There are different methods to accomplish this task; see e.g. \cite{shevchenko2006structural,lambrigger2007quantification} or for a book length treatment \cite{cruz2015fundamental}. In this paper we focus on the estimation of frequency and severity distributions using historical data because we have access to the leading industry dataset from Advisen that contains a substantial number of cyber loss events.
\end{remark}

As mentioned, one popular choice in cyber risk modelling is to utilise the framework of EVT based on the POT method. Under the assumption that the true distribution of $\widetilde{Y}_i$, $i=1,\dots,N$ lies in the domain of attraction of the Generalized Extreme Value (GEV) distribution, loss exceedances over a high enough threshold $u$, i.e. $Y= \widetilde{Y}-u$ conditional on $\widetilde{Y}>u$, can be assumed to follow a Generalized Pareto Distribution (GPD) loss model with the following density:
\begin{equation}
\label{eq:pareto_density}
    g(y;\mu,\tau)=\frac{\tau}{\mu}\left(1+\frac{y}{\mu} \right)^{-(1+\tau)},
\end{equation}
for $y>0$ if $\tau>0$ and $y\in[0,-\mu]$ if $\tau<0$ \cite[see, e.g.] [] {ganegoda2013,chavez2016}.
If $\tau>0$, which is typically the case for financial and actuarial applications, in order to assume that losses can be described by the density given in Equation (\ref{eq:pareto_density}), the distribution function of $\widetilde{Y}$, $F_{\widetilde{Y}}$, must satisfy the following regular variation condition:
\begin{equation}
\label{eq:varying}
    \overline{F}_{\widetilde{Y}}(x) = 1-F_{\widetilde{Y}}(x) \sim x^{-\tau}L(x),\;x\rightarrow \infty
\end{equation}
for some measurable, slowly varying function $L:(0,\infty)\to(0,\infty)$; see, \cite{balkema1974,pickands1975}. According to the asymptotic representation in Equation  (\ref{eq:varying}), it can be shown that if $\tau\in(0,1)$ then $F_{\widetilde{Y}}$ does not have finite moments, while if $\tau\geq 1$ then $F_{\widetilde{Y}}$ has at least a finite first moment. 

Whilst the LDA framework can provide insight into the modelling of the loss process in different classes of cyber risk, this model in its traditional form is incapable of providing a causal link between the leading drivers or risk factors that induce the loss process to occur. To achieve this further level of understanding of cyber risk loss process modelling one must introduce a regression structure into the LDA model components. Furthermore, in cyber risk loss modelling, the loss processes may vary significantly in their statistical attributes over time and over different companies, sectors or cyber risk event types. Therefore, while the LDA can be helpful in identifying the most appropriate distribution for the frequency and severity of cyber events, it might not be flexible enough to capture the nature of cyber risk to the full extend required to address the types of insurance questions posed at the onset of this manuscript. 

Cyber risk events are heterogeneous in nature and can affect a wide variety of industries at varying degrees of penetration and with varying degrees of financial impact. Loss events or the process that leads to the loss can also arise from direct and indirect causes. This can result in many losses from a collection of connected events to the actual IT focused cyber attack. Therefore, to quantify and understand the risk drivers that affect cyber risk loss processes, a methodology allowing to identify the impact of explanatory variables on the frequency and severity distributions will be highly beneficial. The GAMLSS regression framework allows parameters in both the severity and frequency distributions to depend on covariates and, at the same time, to have a more flexible scaling structure than OLS based regression techniques \citep{stasinopoulos2007generalized,rigby2005,ganegoda2013}. Given that the risk from different types of cyber threats is likely to vary over time, we consider the extension to dynamic EVT in \cite{chavez2016}, allowing the parameters to depend also on time, and thus to investigate any non stationary behavior, with particular interest in the tail parameter $\tau$. For any given set of covariates $X$ and time $t$, we consider the following  link functions for the intensity of the Poisson distribution of the loss exceedances $\lambda (X,t)$, the scale parameter of the loss distribution $\mu(X,t)$ and the tail of the loss distribution $\tau(X,t)$ that characterise the LDA loss process:
\begin{align}
&\log(\lambda (X,t)) = f_\lambda(X)+h_\lambda(t),\notag\\
& \log(\mu(X,t)) = f_\mu(X) + h_\mu(t),\label{eq:link}\\
& \log(\tau(X,t)) = f_\tau(X)+h_\tau(t),\notag
\end{align}
where $h_\lambda,h_\mu,h_\tau$ are measurable, twice differentiable functions. The functional form of $f_\lambda,f_\mu$ and $f_\tau$ is assumed to be linear (linear predictors under a log-link).

In the estimation routine, we consider data at the company level, aggregated by year. Let $n_i^c$ and $y_i^c$, be the number of cyber events and  related loss in year $i$ in company $c$, respectively, for $i=1,\dots,K$ and $c=1,\dots,C$.  We aim to estimate the coefficients for the linear predictors that are connected to the parameters of the frequency and severity distribution via the link functions (\ref{eq:link}). Following \cite{chavez2016}, the estimation can be conducted via two penalized maximum likelihood optimization objectives as outlined below in Equation (\ref{Eqn:Freq}) for the frequency and Equation (\ref{Eqn:Sev}) for the severity:
\begin{equation}\label{Eqn:Freq}
\max \sum_{c=1}^C\sum_{i=1}^K\log(f_N(n_i^C;\lambda))-\gamma_\lambda\int_0^Kh''_\lambda(t)dt
\end{equation}
and 
\begin{equation} \label{Eqn:Sev}
\max \sum_{c=1}^C\sum_{i=1}^K\log(g(y_i^C;\mu,\tau))-\gamma_\mu\int_0^Kh''_\mu(t)dt-\gamma_\tau\int_0^Kh''_\tau(t)dt,
\end{equation}
where $f_N(\cdot;\lambda)$ is the probability density function of a Poisson random variable, and   $\gamma_\lambda, \gamma_\mu,\gamma_\tau$ are the smoothing parameters. We adopt the algorithm proposed by Rigby and Stasinopoulos to solve the optimization problems and find the parameter estimates \cite[see,][]{stasinopoulos2007generalized,rigby2005, stasinopoulos2017}. When $h_\lambda,h_\mu,h_\tau$ are chosen to be cubic splines, the penalised maximum likelihood used in the \emph{gamlss} R-package coincides with the one in the \emph{QRM} R-package \cite[see,][]{stasinopoulos2008instructions,chavez2016}.

\section{Cyber Losses Data Description} \label{sec:data}
\begin{figure}[h!]
	\centering
	\subfloat[Number of Losses][{Number of losses resulting in disclosed losses during the period 2008-2020 in a sliding window of 6 months }]{{\includegraphics[scale=0.5]{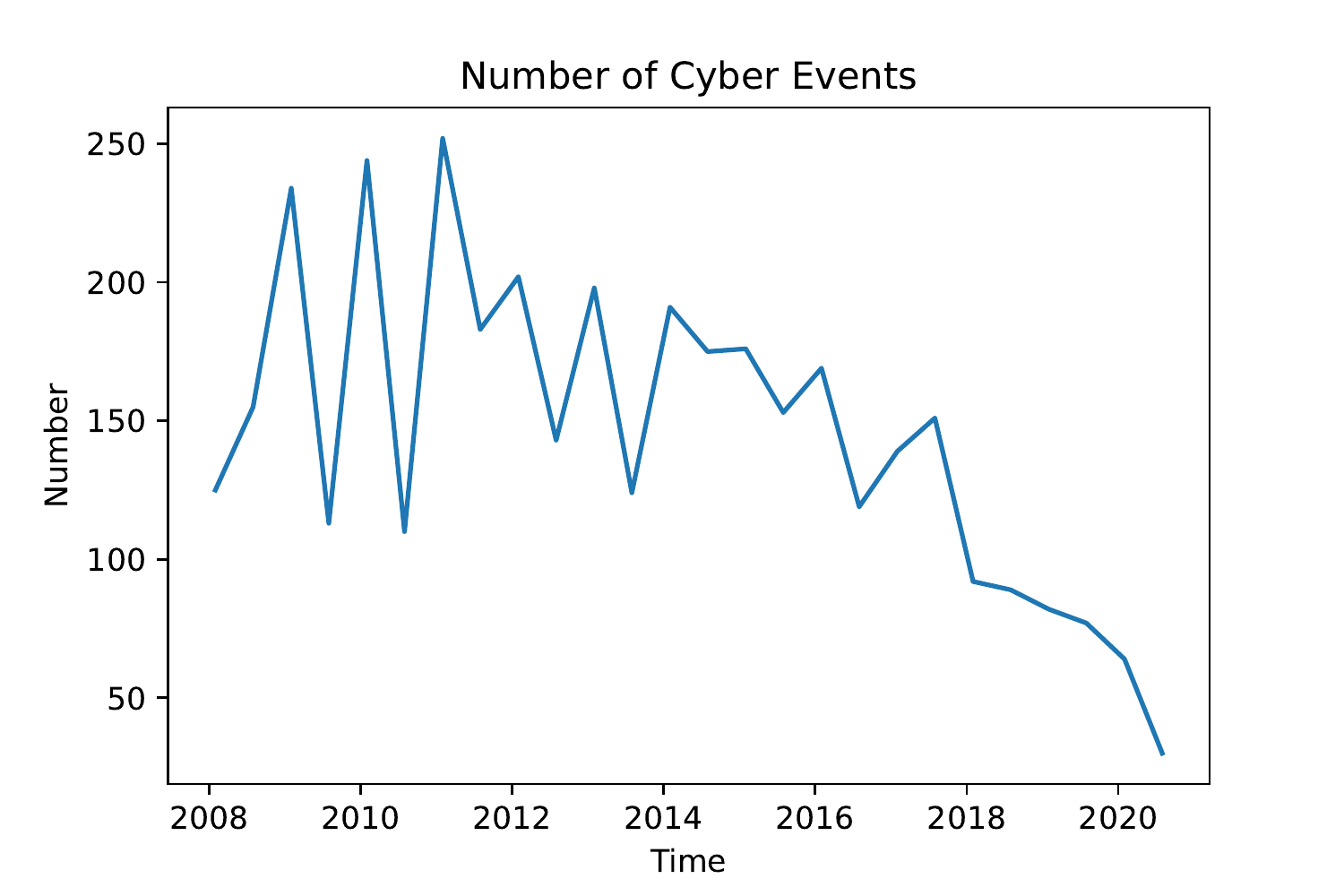}}}
	\qquad
	\subfloat[Mean Loss][{Mean loss during the period 2008-2020 in a sliding window of 6 months}]{{\includegraphics[scale=0.5]{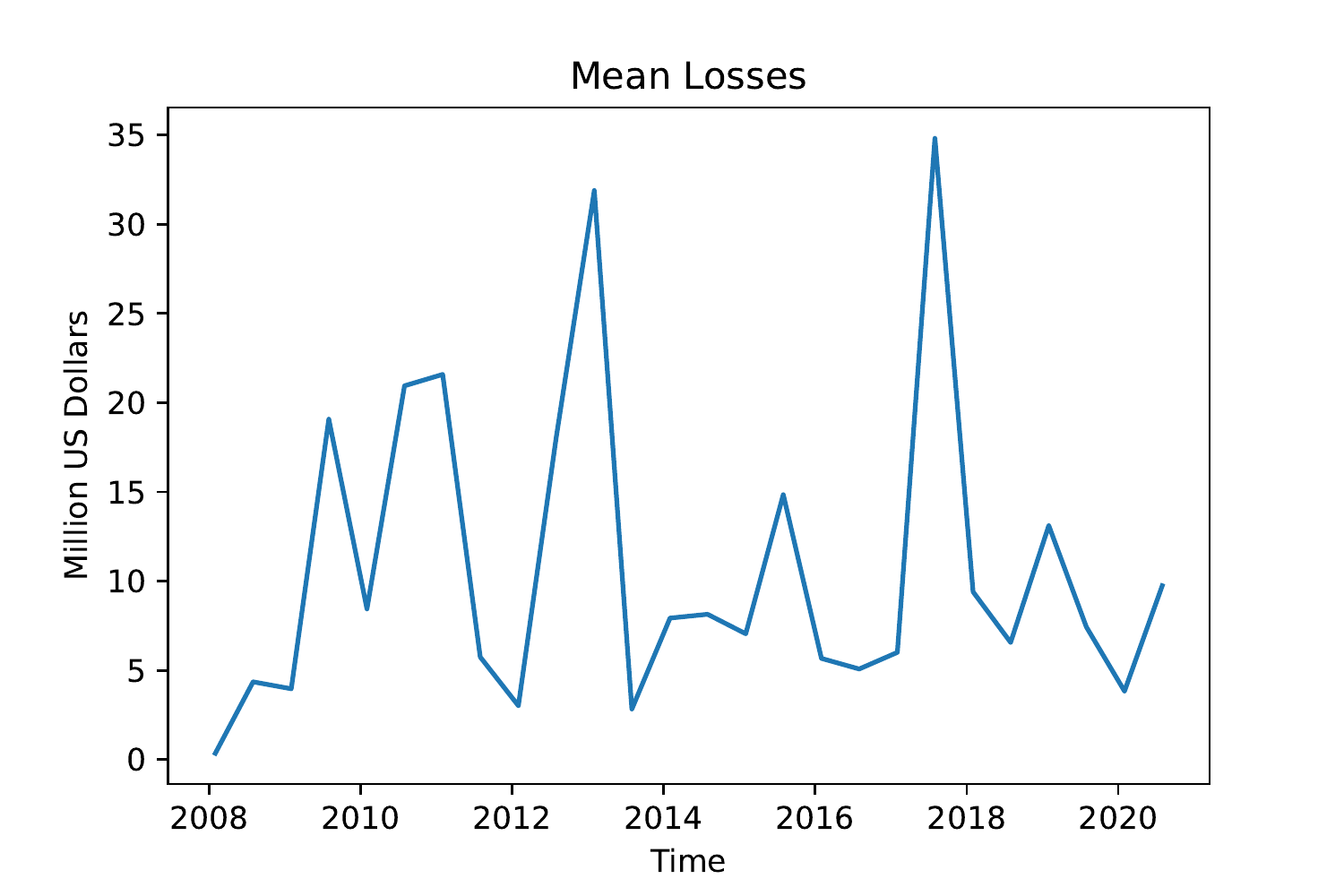}}} 
	\qquad
	\subfloat[Median Loss][{Median loss during the period 2008-2020 in a sliding window of 6 months}]{{\includegraphics[scale=0.5]{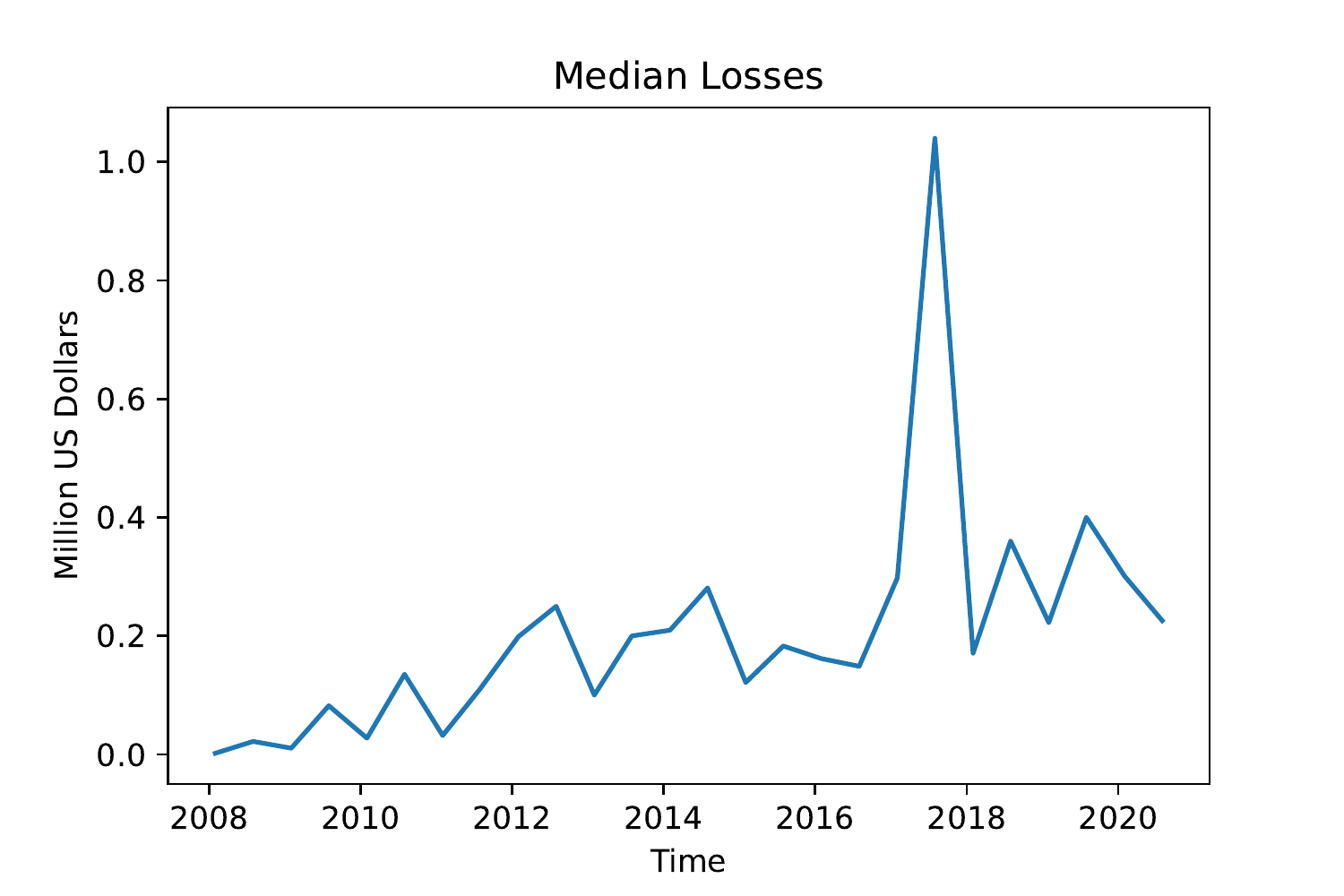}}}
	\caption[Time evolution of losses]{These figures show the time evolution of the number of cyber events and resulting losses from 31 January 2008 to 31 July 2020 in a sliding window of 6 months.}
	\label{Fig:losses_evolution}
\end{figure}

Our comprehensive data set is sourced from Advisen and contains more than 132,126 cyber events from 2008 to 2020, affecting 49,495 organisations across the world\footnote{Given the nature of cyber risk, it can be expected than a great number of events are not recorded in the dataset, since companies are reluctant to report cyber risk related events to the public in order to avoid, among other things, a loss of reputation and trust from their counterparties.}. Advisen is a USA-based for-profit organisation which collects and processes cyber reports from reliable and publicly verifiable sources such as news media, governmental and regulatory sources, state data breach notification sites, and third-party vendors. Given that the interest in cyber risk is on the rise, some recent studies on cyber risk have also made use of Advisen Cyber Loss Data \cite[see, e.g.][]{romanosky2016,aldasoro2020, cyentia2020}. More than 80\% of the events recorded affect organisations residing in the USA and for each event accident timeline (i.e. first notice date, accident date, loss start date, and loss end date), a detailed explanation of the event is provided. One of the key advantages in comparison to other commonly used datasets -- such as e.g. the ``Chronology of Data Breaches" provided by the Privacy Rights Clearinghouse (PRC)\footnote{\url{https://privacyrights.org/data-breaches}} -- is that the Advisen dataset offers direct information of monetary losses linked to each cyber risk event, providing an empirical measurement of financial losses that can be used for modelling purposes. Advisen also provides, when available, cost, economic losses, litigated loss amounts, fines, and penalties for cyber risk events from regulatory bodies. Although the dataset comprises more than 130,000 cyber risk related events, the nature of cyber risk itself determines that only a small percentage of the observation can be used for modelling. Following \cite{edwards2016} and \cite{eling2017data} we remove all observations that do not give information on the monetary losses, and restrict the analysis to the observation for which complete information on company specific characteristics, such as yearly revenue and number of employees are available. This leaves us with a total number of 3,792 observations, corresponding to roughly 2.6\% of the total events. 
\begin{remark}Cyber risk severity can be expressed as number of records or monetary losses. The type of relationship between the two is still the object of ongoing research and debate. According to a recent study by the Ponemon Institute in partnership with IBM security, the logarithm of number of records and the logarithms of cyber risk related monetary losses are linked via a linear relationship\citep{eling2017data,ponemon2019}. Given that one of the objectives of our study is to quantify cyber risk in terms of monetary losses to provide insights to decision makers, practitioners, and insurance market participants, we focus on monetary losses.
\end{remark}
Figure~\ref{Fig:losses_evolution} shows the time evolution of the number of events, as well as the mean and median losses from 31 January 2008 to 31 July 2020 in a sliding window of 6 months. One can readily observe that the number of cyber events increases between 2008 and 2014, and then it seems to follow a decreasing trend contrary to \cite{maillart2010} and \cite{eling2019actual}, who observe a decreasing trend in mean and median of cyber event related losses. We also find that both mean and median follow an increasing trend jointly with an increase in variability. It is also worth to notice that the median is significantly lower than the mean, since the majority of cyber events related losses are of small magnitude and relatively few, very extreme events are recorded. This is consistent with a heavy tailed loss generating process with rarely occurring but severe loss consequences for major cyber events.

The decreasing trend in the number of events observed after 2016 needs to be interpreted in presence of a reporting delay that can be attributed to various factors. It is well known that businesses have a low tendency of reporting cyber related events to avoid repercussions on their reputation. Moreover, Advisen collects data under the USA Freedom of Information Act regulation which prescribes a 60 days window between discovering the data breach and reporting it to affected parties, while non‐USA domiciled entities do not have such strict requirements.

Typically, in OpRisk, loss events are classified according to business line and event type. 
On the one hand, Basel II defines seven official event types and eight business lines for Basel II banking OpRisk practices; see \cite{BCBS2006}. On the other hand, given the multidimensional heterogeneity that characterizes cyber risk, classifying cyber risk is a non-unique task to the point that a world wide accepted cyber risk classification does not exist to date as no consensus on standardisation has been achieved yet; see further discussion in \cite{peters2018understanding,peters2018statistical}. Advisen provides its own classification based on the type of cyber threat\footnote{For business line classification, given that the vast majority of the events affect companies residing in the USA, we adopt the North American Industry Classification System (NAICS), which comprises of 23 business sectors with corresponding codes available in the Advisen dataset.}:
\begin{itemize}
 \setlength\itemsep{0em}
    \item \textbf{Privacy – Unauthorized Contact or Disclosure}: cases when personal information is used in an unauthorized manner to contact or publicize information regarding an individual or an organisation without their explicit permission.
    
    \item \textbf{Privacy – Unauthorized Data Collection}: cases where information about the users of electronic services, such as social media, cellphones, websites, and similar is captured and stored without their knowledge or consent, or where prohibited information may have been collected with or without their consent.
    
    \item \textbf{Data – Physically Lost or Stolen}: situations where personal confidential information or digital assets have been stored on, or may have been stored on, computer, peripheral equipment, data storage, or printouts which has been lost,stolen, or improperly disposed of.
    
    \item \textbf{Data – Malicious Breach}: situations where personal confidential information or digital assets either have been or may have been exposed or stolen, by unauthorized internal or external actors whose intent appears to have been the acquisition of such information.

    \item \textbf{Data – Unintentional Disclosure}: situations where personal confidential information or digital assets have either been exposed, or may have been exposed, to unauthorized viewers due to an unintentional or inadvertent accident or error.
    
    \item \textbf{Identity – Fraudulent Use/Account Access}: identity theft or the fraudulent use of confidential personal information or account access in order to steal money, establish credit, or access account information, either through electronic or other means.
    
    \item \textbf{Industrial Controls and Operations}: losses involving disruption or attempted disruption to ``connected" physical assets such as factories, automobiles, power plants, electrical grids, and similar (including “the internet of things”).
    
    \item \textbf{Network/Website Disruption}: unauthorized use of or access to a computer or network, or interference with the operation of same, including virus, worm, malware, digital denial of service (DDOS), system intrusions, and similar.

    \item \textbf{Phishing, Spoofing, Social Engineering}: attempts to get individuals to voluntarily provide information which could then be used illicitly, e.g. phishing or spoofing a legitimate website with a close replica to obtain account information, or sending fraudulent emails to initiate unauthorized activities (aka “spear phishing”).
    
    \item \textbf{Skimming, Physical Tampering}: use of physical devices to illegally capture electronic information such as bank account or credit card numbers for individual transactions, or installing software on such point-of-sale devices to accomplish the same goal.
    
    \item \textbf{IT – Configuration/Implementation Errors}: losses resulting from errors or mistakes which are made in maintaining, upgrading, replacing, or operating the hardware and software IT infrastructure of an organisation, typically resulting in system, network, or web outages or disruptions.
    
    \item \textbf{IT – Processing Errors}: losses resulting from internal errors in electronically processing orders, purchases, registrations, and similar, usually due to a security or authorization inadequacy, software bug, hardware malfunction, or user error.
    
    \item \textbf{Cyber Extortion}:  threats to lock access to devices or files, fraudulently transfer funds, destroy data, interfere with the operation of a system/network/site, or disclose confidential digital information such as identities of customers/employees, unless payments are made.
    
    \item \textbf{Undetermined/Other}.
\end{itemize}

Other possible classifications of cyber risk events are available in the literature. For example, \cite{eling2019actual} suggests to divide cyber risk events into four categories, according to OpRisk classification: ``Actions by People", ``System and Technical Failure", ``Failed Internal Process", ``External Events" \cite[see,][]{cebula2010,cebula2014}. On the other hand, \cite{romanosky2016} provides cyber risk driven categories, such as ``Data Breach",	``Security Incident",	``Privacy Violation",	``Phishing Skimming", and ``Other".   

Table~\ref{table:descriptive} provides descriptive statistics of non-zero losses for each cyber risk category in the Advisen dataset. We find that the specified cyber risk classification exhibits great variability in terms of the number of recorded events in each risk category. This indicates that certain loss events are more prevalent, however this changes over time in an inhomogeneous manner. Furthermore, there is a heterogeneity in the severity distributions as evidenced by the first four moments of the loss distributions for each risk category. All risk categories exhibit a mean loss that is higher than the median, indicating skewness and potential for heavy tails and leptokurtic behaviour. In some cases this effect is so pronounced that the mean is two orders of magnitude higher than the median and paired with high kurtosis. Additional goodness of fit analysis can be found in Section~\ref{appendix:ks} of the Appendix. 

\begin{table}[h]
	\centering
	\caption[Descriptive Statistics]{This table reports descriptive statistics of cyber risk related losses aggregated by categories. All dollar values are reported in million dollars. The losses exhibit great variability in terms of median and first four moments across the considered risk types. ``IT  –  Configuration/Implementation  Error", ``Privacy – Unauthorized Data Collection", and ``Industrial Controls" have the highest average loss amongst all cyber risk categories.}
	\label{table:descriptive}
	\begin{tabular}{l|rrrrrr}
		Risk category                        &	N	& Mean	&  Median &  StDev &  Skew	 & Kurt \\
		\hline
		
		Privacy - Unauthorized Contact or Disclosure	          & 1417 & 3.56   &	0.03  &	27.33     & 27.33  &	919.39\\
		Privacy – Unauthorized Data Collection 	          & 113  & 54.69  &	0.45  &	472.56	  & 10.15  &	103.14\\
		Data - Physically Lost or Stolen          & 94	& 24.62  &	0.24  &	206.4     & 9.33   &	86.26\\
		Identity – Fraudulent Use/Account Access		          & 624  & 1.10   &	0.03  &	6.56	  & 10.96  &	136.41\\
		Data - Malicious Breach			          & 719  & 24.94  &	0.53  &	187.02	  & 15.94  &	303.05\\
		Phishing, Spoofing, Social Engineering 			          & 179  & 8.91   &	0.55  &	54.28	  & 12.11  &	153.21\\
		IT -  Configuration/Implementation Errors & 57 	& 18.76  &	0.82  &	45.58	  & 2.95   &	8.61\\
		Data - Unintentional Disclosure           & 175  & 1.52   &	0.12  &	9.74	  & 11.59  &	141.28\\
		Cyber Extortion				              & 110  & 0.63   &	0.01  &	3.11	  & 6.03   &	37.13\\
		Network/Website Disruption		          & 159  & 21.17  &	0.18  &	73.35	  & 4.39   &	19.67\\
		Skimming, Physical Tampering:	          & 84	& 1.85   &	0.05  &	6.35	  & 5.78   &	38.07\\
		IT – Processing Errors 	  & 39	& 86.36 &	2.60  &	283.38	  & 4.80   &	24.06\\
		Industrial Controls 			          & 6	& 30.69  &	2.07  &	68.35	  & 1.35   &	-0.1\\
		Undetermined/Other				          & 16	& 1.74   &	0.62  & 2.8       & 2.61   &	6.38\\

	 	\hline	
		
	\end{tabular}
\end{table}

The descriptive statistics in Table~\ref{table:descriptive} seem to be consistent with previous findings in the literature, that cyber risk related losses follow heavy tailed distributions; see, e.g. \cite{maillart2010,edwards2016,eling2017data}. One of the main objectives of the approach described in Section~\ref{sec:model}, is to estimate the tail parameter $\tau$ using a parametric approach to draw inference on how specific covariates can impact on the risk profile of cyber risk events. Here, we also present a non parametric estimate for the tail parameter, the Hill's estimator; see, e.g. \cite{hill1975,grama2008,durrieu2015}, i.e. for a given threshold $k$, we consider the following estimator:
\begin{equation}
\label{e:hill}
1/\widehat \tau_k(n)=\xi_k(n):= \frac{1}{k} \sum_{i=1}^{k} \log \Bigg(\frac{y_{(n-i+1,n)}}{y_{(n-k,n)}} \Bigg),
\end{equation}
where $1\leq k\leq n-1$ and $y_{(n,n)} \geq y_{(n-1,n)}\geq  \cdots \geq  y_{(1,n)}$ are the order statistics of the sample $y_i,\ i=1,\ldots,n$.
\begin{figure}[ht]
\begin{center}
\includegraphics[width=\textwidth, height=8.5cm]{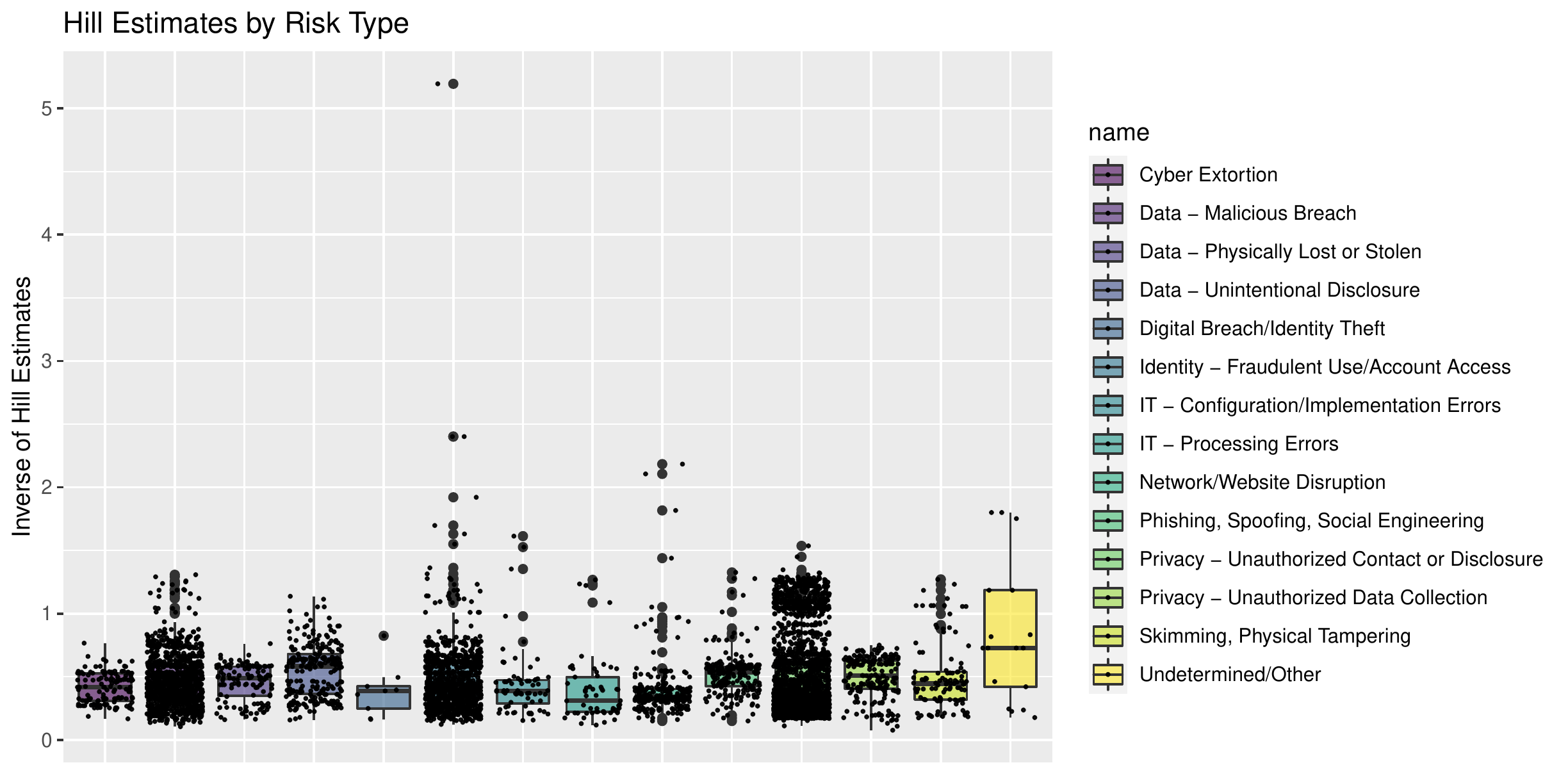}
\end{center}
\caption{Hill's estimates of monetary losses linked to cyber events from 2008 to 2020 by risk types. }
\label{Fig:hill_estimator}
\end{figure}

\begin{figure}[ht]
\begin{center}
\includegraphics[width=\textwidth, height=8.5cm]{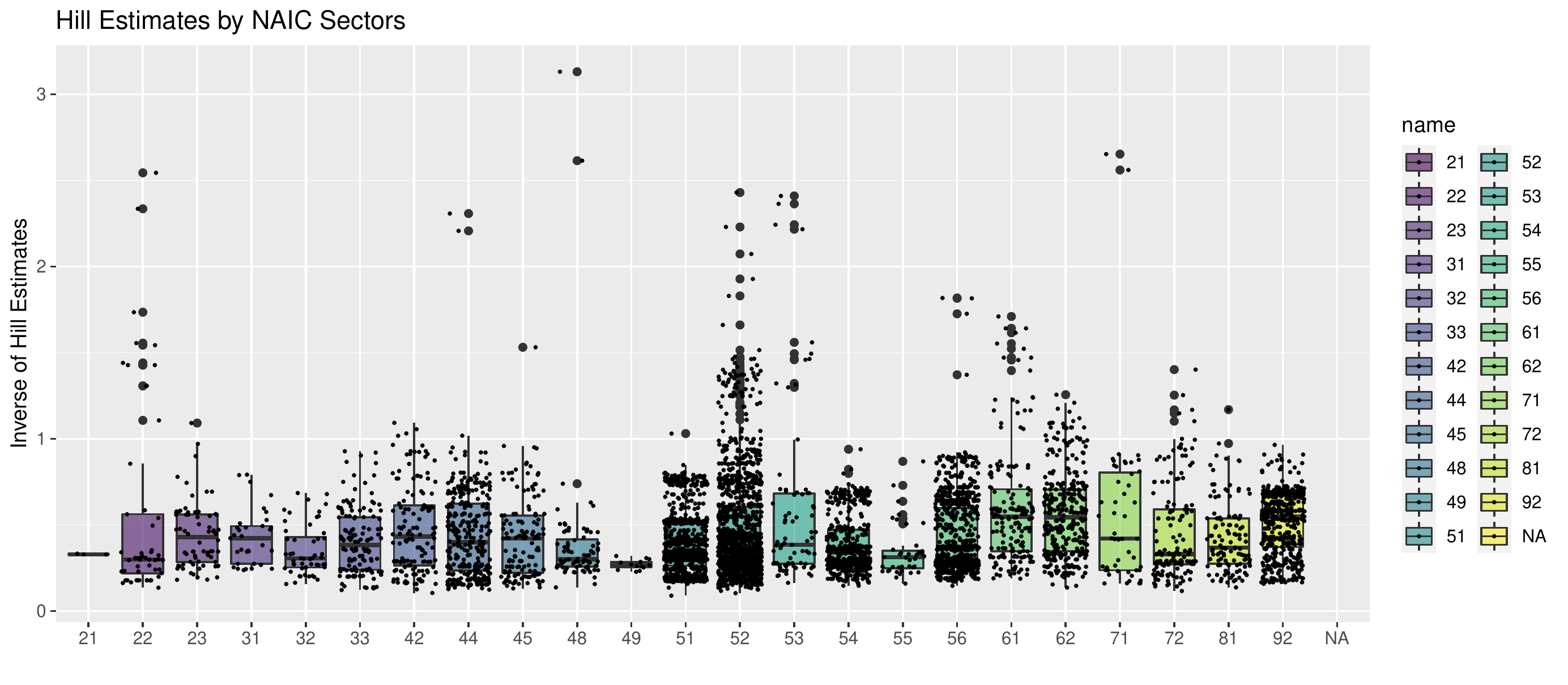}
\end{center}
\caption{Hill's estimates of monetary losses linked to cyber events from 2008 to 2020 by business sectors.}
\label{Fig:hill_estimatorBS}
\end{figure}

Figures~\ref{Fig:hill_estimator} and \ref{Fig:hill_estimatorBS} show the values for $\widehat{\tau}(k)$ broken down by cyber risk types and business sectors in the North American Industry Classification System (NAICS)\footnote{\url{https://www.census.gov/naics/}}, for different threshold values of the level in the Hill estimators order statistics used in estimator in Equation (\ref{e:hill}). As demonstrated in the box plots of the estimates for each risk category (Figure~\ref{Fig:hill_estimator}) and NAICS business sector (Figure \ref{Fig:hill_estimatorBS}), both decompositions demonstrate the typical behaviour of the sample estimators for the population tail, i.e. both the median and the interquartile range (IQR) of the tail estimator for the companies involved in the aggregation suggest a tail index estimate lower than 0.5. This implies that the loss distributions exhibit extreme heavy tails and do not have finite first moments. There are some risk categories, where this behaviour is less pronounced such as in ``Privacy - Unauthorised Contact or Disclosure" where whilst the median and IQR of companies affected by this risk type have median and IQR consistent with infinite mean heavy tailed models, there is however a wide variety of variation in this case compared to other risk categories. We also suggest not to put too much emphasis on the results for the category ``Undetermined/Other" due to its catch-all effect for many sources of loss events that cannot be allocated to any of the other 14 categories. 

The key takeaway from this preliminary analysis is that, by and large, cyber related loss distributions are of the infinite mean type, regardless of the business sectors or risk category. The implications of these empirical findings are twofold. First, they justify the use of the POT method described in Section~\ref{sec:model} for estimating the tail parameter, rather than focusing on OLS regression techniques, given that the data itself exhibits heavy tails. Second, the degree of heavy-tailedness of the cyber risk severity distribution poses a key aspects of our analysis. It could be argued that the parameter $\tau$ in Section~\ref{sec:model} is one of the main factors discriminating the insurability of cyber risk. Having observed tail values of the Hill estimates lower than 1, this behaviour of losses could jeopardize cyber risk insurability. 

\FloatBarrier

\section{GAMLSS Regression Analysis of Cyber Risk Losses} 
\label{Sec_gamlss_analysis}
In this section, we present the empirical results of the applied GAMLSS regression to further examine the Advisen Cyber Loss Data. First, we discuss which covariates should be included in the analysis of the frequency and severity model parameters according to the literature. Second, we analyse whether cyber risk types should be modelled jointly or separately. Finally, we present the results of the POT method combined with the GAMLSS approach.

\subsection{GAMLSS Regression Covariates}
The GAMLSS approach allows us to identify relevant risk drivers and covariates as well as their impact on distributional parameters of the frequency and severity distribution in the LDA framework. The choice of covariates is dictated by problem specific instances, available information, as we all as previous findings in the literature on drivers of cyber risk. Several studies have observed changes in loss frequency and severity over time \citep{maillart2010, biener2015insurability, chavez2016, eling2019actual}. Observed changes in the number of events as well as in the magnitude of losses in Figure~\ref{Fig:losses_evolution} also seems to confirm the time-varying nature of cyber risk that has been observed by these authors for other datasets. Consequently, time is included as a covariate in the GAMLSS regression framework for both frequency and severity. It is well documented in the literature that company size positively correlates with loss severity \cite[see, among others,][]{cope2008,ganegoda2013, eling2019actual}. Thus, we develop a proxy for company size using revenue and the number of employees at the time of the cyber event. Controlling for the number of employees should also address the fact that human behavior is one of the major drivers for cyber risk as it has often been reported in the IT literature \citep{evans2016}. Another typical covariate, having a relationship with loss severity and frequency is the industry sector; see \cite{dahen2010,ganegoda2013,chavez2016,eling2019actual}. Since the majority of cyber events reported by Advisen are located in the USA and therefore regulated by the Freedom of Information Act (FOIA) and Health Information Technology for Economic and Clinical Health Act (HITECH), we distinguish between Financial industry, Healthcare industry, and other\footnote{Under the FOIA and HITECH, the Secretary for Health and Human Services has to publicly disclose breaches of unsecured protected health information affecting 500 or more individuals. For more information, see  https://www.hhs.gov/.}.  Following \cite{eling2019actual} we consider two additional covariates: \emph{geographical location} and \emph{contagion}. Location accounts for regulatory differences between different regions, i.e. in the USA-focused setting of this study for regulatory differences between the USA and the rest of the world. Given that the majority of cyber risk events are in the USA, we distinguish between companies residing in the USA and the rest of the world. Contagion can be classified as three mutually exclusive realizations: events related to multiple losses in a single company, events related to multiple losses in different companies, and one shot events. Similarly, it is also common practice to include event categories as covariates to control for differences in the relationship for NAICS industry sectors or cyber risk loss categories.

Given that many covariates are company or institution specific we look at loss frequency and severity at the firm level. Therefore for each event we consider the following series of categorical variables: 
\begin{itemize}
\setlength\itemsep{0em}
\item $RT_s$, $s=1,\dots, S$ indicating the cyber risk type; 
\item $L_l$, $l\in\mathcal{L}=\{\text{USA, non-USA}\}$, indicating if the company that was affected by the cyber event resides in the USA or not;
\item $B_b$, $b\in\mathcal{B}=\{\text{Finance, Healthcare, Other}\}$, indicating the company business sector; 
\item $R_r$, $r\in\mathcal{R}=\{\text{small,medium,big}\}$  indicating three classes of company revenue (small, if the revenue is lower than the 33rd percentile, medium if the company revenue is between the 33rd and 66th percentiles, and big if the company revenue is higher than the 66th percentile); 
\item $E_e$, $e\in\mathcal{E}=\{\text{small,medium,big}\}$  indicating three classes of company size with respect to the number of employees (small, if the number of employees is lower than the 33rd percentile, medium if the number of employees is between the 33rd and 66th percentiles, and big if the number of employees is higher than the 66th percentile). 
\end{itemize}
Furthermore, we also include two additional dummy variables, $ML$ and $MC$, indicating whether the event causes multiple losses in a single company or in various companies, respectively.

\subsection{GAMLSS Model Selection}
\label{SubSec:JointDecoupled}
In this section we explore the ongoing debate regarding the effect of modelling cyber risk frequency and severity jointly across risk types or as individual models for each risk type within the LDA model framework \cite[see, e.g. discussion in][]{edwards2016,eling2017data,eling2018copula,eling2019actual,jung2021}. To explore this issue, we combine the POT method of EVT for severity within the regression framework of the GAMLSS approach. We compare a joint estimation of the GAMLSS regression structure across all risk types of the Advisen dataset against a separate estimation of a GAMLSS for the frequency and severity components of each risk type. Thus, we will be able to utilise formal hypothesis testing based on likelihood inference procedures to seek evidence for or against these approaches.

We distinguish the coupled model estimation approach as follows (termed the joint model), where we consider the following likelihood and link functions:
\begin{align}
\label{eq:joint_model}
    & \log\left(L_1(Y;\mu,\tau)\right) =\sum_{i=1}^m \log\left(g(y_i;\mu,\tau)\right),\\ 
    & \log\left(\mu(X,t)\right) = \beta_0+\sum_{s=1}^{S} \beta_s RT_s+\sum_{l\in \mathcal{L}} \beta_lL_l+\sum_{b\in \mathcal{B}}\beta_bB_b+\sum_{r\in\mathcal{R}} \beta_rR_r+\sum_{e\in \mathcal{E}}\beta_eE_e\notag\\
    &\hspace{2.6cm} +\beta_{ml}ML+\beta_{mc}MC+\beta_tt,\notag\\
    & \log\left(\tau(X,t)\right) = \beta_0+\sum_{s=1}^{S} \beta_s RT_s+\sum_{l\in \mathcal{L}} \beta_lL_l+\sum_{b\in \mathcal{B}}\beta_bB_b+\sum_{r\in\mathcal{R}} \beta_rR_r+\sum_{e\in \mathcal{E}}\beta_eE_e\notag\\
    &\hspace{2.6cm}+\beta_{ml}ML+\beta_{mc}MC+\beta_tt,\notag
\end{align}
and the decoupled model (termed the separate risk category model) as follows:
\begin{align}
   & \log\left(L_2(Y;(\mu_s)_{s=1,...,S},(\tau_s)_{s=1,...,S})\right) = \sum_{s=1}^S\sum_{i=1}^{n_s}\log(g(y_i^{(s)};\mu_s,\tau_s)), \notag\\
     & \log\left(\mu_s(X,t)\right) = \beta_0+\sum_{l\in \mathcal{L}} \beta_lL_l+\sum_{b\in \mathcal{B}}\beta_bB_b+\sum_{r\in\mathcal{R}} \beta_rR_r+\sum_{e\in \mathcal{E}}\beta_eE_e+\beta_{ml}ML+\beta_{mc}MC+\beta_tt,\notag\\
    & \log\left(\tau_s(X,t)\right) = \beta_0+\sum_{l\in \mathcal{L}} \beta_lL_l+\sum_{b\in \mathcal{B}}\beta_bB_b+\sum_{r\in\mathcal{R}} \beta_rR_r+\sum_{e\in \mathcal{E}}\beta_eE_e+\beta_{ml}ML+\beta_{mc}MC+\beta_tt,\notag
\end{align}
where each coefficient $\beta$ refers only to the equation it appears in and additional subscripts referring to parameter or risk type are omitted for notational convenience. We apply Vuong's closeness test \citep{vuong1989} to determine if one can statistically distinguish between the joint and decoupled modelling approach. The test is essentially a likelihood-ratio test for model selection based on the Kullback–Leibler information criterion. To conduct the test, in a first step we formulate the null and alternative hypothesis as follows: 
\begin{equation*}
    H_0:\:\mathrm{var}\left(\log\left(\frac{L_1(Y;.,.)}{L_2(Y;.,.)}\right)\right) = 0 \qquad \mbox{and}\quad \: H_1:\:\mathrm{var}\left(\log\left(\frac{L_1(Y;.,.)}{L_2(Y;.,.)}\right)\right) \neq 0.
\end{equation*}

Based on the null hypothesis, we then aim to assess whether the two models are equally close to the true data generating process, against the alternative that one model is closer. Thus, in our setup we aim to verify if the following condition is satisfied based on the estimates for the joint and decoupled GAMLSS model structure:
\begin{equation}
    g(y,\widehat{\mu},\widehat{\tau})=g(y,\widehat{\mu}_s,\widehat{\tau}_s),\qquad s=1,\dots,S,\notag
\end{equation}
where $\widehat{\mu}$ and $\widehat{\tau}$ are the fitted parameters of the joint model, and $\widehat{\mu}_s$ and $\widehat{\tau}_s$ are the fitted parameters under the decoupled model in which the model is fitted separately for each risk type. 

The test statistic under the null hypothesis is given by the variance of the likelihood ratio and, under suitable regularity conditions, follows a $\chi^2$ distribution with degrees of freedom equal to the number of parameters.

Applying the test to the proposed GAMLSS structure, our results suggest that we do not have enough evidence to reject the null hypothesis and therefore, the two alternatives appear to be indistinguishable. This last point further confirms the findings of the previous section and supports the statement that cyber event related losses are so heavy tailed that it is not possible to distinguish between the two nested models. In other words across all Advisen risk type categories, one cannot statistically distinguish between the loss types, at least in their important attributes from an insurance perspective, relating to the tail behaviour of cyber loss processes for each risk category. As a consequence, we are confident to therefore proceed the analysis by looking at separate models for each risk category given that the estimation routines require less computing time.

Table~\ref{table:stats_threshold} shows threshold values and descriptive statistics for each category; ``Privacy - Unauthorized Contact or Disclosure" has the highest threshold, both in percentage and in millions, while for the rest of the risk types the required threshold is almost negligible.
\begin{table}[h]
	\centering
	\caption[Descriptive Statistics Threshold]{This table reports descriptive statistics of cyber risk related losses that exceed the corresponding threshold values for each risk category. All dollar values are reported in million dollars. The threshold for each risk category is chosen as the lowest possible value ensuring that the null hypothesis of GPD cannot be rejected for the exceeding losses.}
	\label{table:stats_threshold}

	\begin{adjustbox}{center,scale=0.8}
    	\begin{tabular}{l|rrrrrrrrr}
    		Risk Type                        & $u$ & Min & N & Mean & Median & StDev & Skewness & Kurtosis\\
    		\hline
    		
    		Privacy - Unauthorized Contact or Disclosure 	          &0.63 &	0.2    & 530 & 9.49	 &  2.92 &	44.08 &	17.58 &	351.81\\
    		Privacy – Unauthorized Data Collection  	          &-    &	0.00005& 113 & 54.69 & 	0.45 &	472.56&	10.15 &	103.14\\
    		Data - Physically Lost or Stolen          &-    &	0.0003& 94  & 24.62 &	0.24 &	206.4 &	9.33  &	86.26\\
    		Identity – Fraudulent Use/Account Access		          &0.4  &	0.0156& 375 & 1.83  &	0.16 &	8.38  &	8.47  &	80.82\\
    		Data - Malicious Breach			          &0.14 &	0.0263 & 619 & 28.96 &	0.94 &	201.29&	14.78 &	261.09\\
    		Phishing, Spoofing, Social Engineering 			          &-    &	0.0008 & 179 & 8.91  &	0.55 &	54.28 &	12.11 &	153.21\\
    		IT - Configuration/Implementation Errors  &-    &	0.0012& 57	 & 18.76 &	0.82 &	45.58 &	2.95  &	8.61\\
    		Data - Unintentional Disclosure           &0.04 &	0.0035& 169 & 1.57  &	0.13 &	9.91  &	11.38 &	136.27\\
    		Cyber Extortion					          &0.02 &	0.0002& 109 & 0.64  &	0.01 &	3.12  &	6	  &36.75\\
    		Network/Website Disruption		          &-    &	0.0006 & 159 & 21.17 &	0.18 &	73.35 &	4.39  &	19.67\\
    		Skimming, Physical Tampering	          &-    &	0.0003& 84	 & 1.85  &	0.05 &	6.35  &	5.78  &	38.07\\
    		IT - Processing Errors			          &-    &	0.0003 & 39	 & 86.36 &	2.6  &	283.38&	4.8   &	24.06\\
    		Industrial Controls and Operation         &0.51 &	1.1348& 4	 & 46.03 &	6.5  &	82.73 &	0.75  &	-1.69\\
    		Undetermined/Other				          &-    &	0.0079& 16	 & 1.74  &	0.62 &	2.8	  & 2.61  &	6.38\\
    
    	 	\hline	
    		
    	\end{tabular}
	\end{adjustbox}

\end{table} 

\subsection{Dynamic Extreme Value Theory Estimates}
In this section we present the results for cyber event frequency and severity regression models, which were estimated based on the assumptions of independence between frequency and severity and as such the estimation can be performed separately. For the frequency of cyber events we consider a Poisson distribution for each risk type, where the intensity parameter depends on covariates according to the following link function, where subscripts referring to the parameter $\lambda$ and risk type are omitted for notational convenience:
\begin{equation}
    \label{eq:link_intensity}
    \log\left(\lambda_s(X,t)\right) = \beta_0+\sum_{l\in \mathcal{L}} \beta_lL_l+\sum_{b\in \mathcal{B}}\beta_bB_b+\sum_{r\in\mathcal{R}} \beta_rR_r+\sum_{e\in \mathcal{E}}\beta_eE_e+\beta_{ml}ML+\beta_{mc}MC+\beta_tt.
\end{equation}
Here we assume $h_{\lambda}$ to be linear in time; a detailed implementation of the spline smoothing approach in \cite{chavez2016} to determine the functional form of $h_{\lambda}$ can be found in Section~\ref{appendix:non_linearity_interaction} of the Appendix.  Table~\ref{table:model0_lambda} reports the score ratios of each covariate for each risk type. Fitting the cyber risk types separately proves to be a flexible approach that allows us to capture the heterogeneity in the nature of losses and their driving factors for each risk category. By looking at the estimates, the impact of time on cyber event frequency clearly depends on the risk types. The frequency of cyber events related to privacy, identity, data breaches, and skimming and tampering is negatively related with time. These risk types compose the vast majority of cyber events in the dataset, which also explains the decreasing trend in Figure~\ref{Fig:losses_evolution}. Even though we find consistent evidence of a decreasing trend in the number of cyber events, confirming the results in \cite{eling2019actual}, we cannot rule out the fact that the frequency of events in the categories ``Cyber Extortion" and ``Phishing and Spoofing" is increasing over time. Moreover, many recent industry and technical reports have indicated that cyber extortion events are becoming more frequent, even though companies tend to report less cyber extortion events than other cyber event types \cite[see, e.g.][]{europol2020,falk2021}. Recall that we measure the impact of company size on the frequency of events by considering the number of employees and revenue of the company. Also for these variables, we find substantial differences between the individual risk categories: identity, phishing, and cyber extortion are more frequent in companies with low average labour productivity (medium and high number of employees and low revenue), indicating that for these risk categories the frequency is higher for companies lacking efficient resource allocation, technological know-how, or returns of scale. On the other hand, companies with medium and high revenue suffer more frequently from privacy related events. The frequency of cyber events also depends on business sectors. Financial and health sectors have similar effects in privacy, ``Data - Physically Lost or Stolen", and ``Phishing, Spoofing, Social Engineering". ``Cyber Extortion", and ``Data - Unintentional Disclosure" happen to be less frequent in the financial sector than in healthcare, due to the strict regulation around data policy in the financial sector. Looking at the effect of contagion, only the frequency of ``Privacy - Unauthorized Contact or Disclosure", ``Data - Physically Lost or Stolen " and ``Data - Malicious Breach" results appear to be impacted by events spilling over from company-to-company.

\begin{table}[h]
    	\centering
        	\caption{This table reports the regression score ratios for parameters in the link function $\log(\lambda)$. The significance of each coefficient is reported with the usual convention  of 1,2, and 3 stars corresponding to a p-value of $10\%$, $5\%$, and $1\%$, respectively.}
        	\label{table:model0_lambda}
       \begin{adjustbox}{center,scale=0.55}
        	\begin{tabular}{lllllllllllll}
        		RiskType & $\beta_0$ & Time        & $R_{medium}$      &	$R_{big}$    &	$E_{medium}$   &	$E_{big}$  &	$L_{USA}$	      & $B_{financial}$       & $B_{health}$        & ML       & MC       \\
        		
        		\hline
                    Privacy - Unauthorized Contact or Disclosure           & 17.8754*** & -18.6083*** & 6.6067***  & 4.7541*** & -4.2399*** & -6.7822*** & 6.5116*** & -4.7865*** & -6.1643*** & -0.0046 & 7.2530*** \\
                    Privacy – Unauthorized Data Collection              & 0.4137     & -2.5524**   & 0.4517     & 3.1365*** & -1.3059    & 0.0869     & 1.4602    & -3.8447*** & -2.4056*** & 0.0514  & 0.0655   \\
                    Data - Physically Lost or Stolen         & 4.6797***  & -4.8496***  & -2.5965*** & -1.3731   & 0.3511     & 1.4979     & -2.9452***& 1.6965*    & 8.3679***  & -0.0547 & 4.5084*** \\
                    Identity – Fraudulent Use/Account Access                & 0.2037     & -15.9455*** & -8.4701*** & -6.3861***& 6.0657***  & 8.0248***  & 7.5698*** & 16.0172*** & -1.4613    & 0.0087  & 0.0123 \\
                    Data - Malicious Breach                  & 4.6599***  & -5.1782***  & -1.03616   & 3.0066*** & 2.9143***  & 6.4114***  & -0.8918   & -0.3428    & 0.3386    & 0        & 8.3027***         \\
                    Phishing, Spoofing, Social Engineering        & -0.4375    & 7.7871***   & -3.9149*** & -1.9394*  & 5.1092***  & 5.3348***  & -1.1372   & -2.5014*** & -1.9951*   & 0        & 0.02266     \\
                    IT – Configuration/Implementation Errors & -0.2305    & 0.6214      & 1.6194*   & 2.6201*** & -0.6968    & 0.1246     & -2.8849***& 0.2772     & 5.1172***  & 0.0430  & 0.0571  \\
                    Data - Unintentional Disclosure          & 0.2291     & -2.1061**   & 0.68469    & 1.7449*   & -1.3631    & -1.2887    & -8.4466***& -1.7927*   & 5.3561***  & 0        & 0.0627        \\ 
                    Cyber Extortion                          & -1.5525    & 7.8595***   & -3.9842*** & -3.4462***& 3.9356***  & 2.8025***  & 0.3568    & -3.3763*** & 0.3533     & 0        & 0.0513        \\
                    Network/Website Disruption               & -0.3285    & 1.1902      & -1.6372*   & -1.5218   & 1.3012     & 4.8584***  & -0.5805   & -3.5876*** & -1.3433    & 0        & 0.0767     \\
                    Skimming, Physical Tampering          & 1.2323     & -4.9652***  & -1.4915    & 0.5161    & -0.0864    & 1.2158     & 2.9849*** & 6.2142***  & -0.1249    & 0.0540  & 0.0699  \\
                    IT - Processing Errors                   & 0.0843     & -0.8353     & -0.6775    & 0.1934    & -1.0146    & 0.7012     & -2.5285***& 3.8826***  & -0.0149    & 0        & 0.0305  \\
                    Industrial Controls and Operations       & -0.1066    & 1.1032      & -0.0091    & -0.193    & 0.0091     & 0.0091     & -0.5626   & -0.0069    & -0.0043    & 0        & 0.0084    \\
                    Undetermined/Other                       & -0.0940    & 0.4228      & 0.1462     & 1.3643    & -0.6335    & -0.725     & 1.1165    & 3.1339***  & -0.0093    & 0        & 0.0196        \\
    		
        		\hline	
        
        	\end{tabular}
     	
         \end{adjustbox}
\end{table}
For cyber event severity we consider a GPD for each risk type, with the following link functions:

\begin{align}
    & \hspace{-0.7cm} \log(\mu_s(X,t)) = \beta_0+\sum_{l\in \mathcal{L}} \beta_lL_l+\sum_{b\in \mathcal{B}}\beta_bB_b+\sum_{r\in\mathcal{R}} \beta_rR_r+\sum_{e\in \mathcal{E}}\beta_eE_e+\beta_{ml}ML+\beta_{mc}MC+\beta_tt,\label{eq:link_severity_scale}\\
    & \hspace{-0.7cm}\log(\tau_s(X,t)) = \beta_0+\sum_{l\in \mathcal{L}} \beta_lL_l+\sum_{b\in \mathcal{B}}\beta_bB_b+\sum_{r\in\mathcal{R}} \beta_rR_r+\sum_{e\in \mathcal{E}}\beta_eE_e+\beta_{ml}ML+\beta_{mc}MC+\beta_tt.\label{eq:link_severity_tail}
\end{align}
Here we assume $h_{\mu}$, and $h_{\tau}$ to be linear in time; a detailed implementation of the spline smoothing approach in \cite{chavez2016} to determine the functional form of $h_{\mu}$, and $h_{\tau}$ can be found in Section~\ref{appendix:non_linearity_interaction} of the Appendix. Tables~\ref{table:model0_mu} and~\ref{table:model0_sigma} show the results for cyber event severity. From an insurance and risk management perspective, one is primarily interested in the tail index relationships. In this section we therefore focus the discussion on the tail parameter $\tau_s$. A tail parameter lower than 1 implies that the distribution has no finite moments. In such a case, the expected loss of the corresponding cyber risk type is infinite. Therefore, in Table~\ref{table:model0_sigma}, covariates with a negative estimated coefficient increase cyber risk type riskiness. Similarly to cyber event frequency, severity of privacy and identity related events follows a decreasing time trend, while ``Data - Malicious Breach", ``Phishing, Spoofing, Social Engineering ", and ``Cyber Extortion" severity has increased over time.

Looking at impact of company size, the picture is less clear for the severity regression models. Losses in the category ``Privacy – Unauthorized Data Collection " are more severe for labour intensive companies, suggesting the relevance of human behavior for these kind of events. Nevertheless, no other privacy type of events supports the same findings. Losses in the category ``Data - Physically Lost or Stolen" are less severe in companies with medium and high number of employees, while ``Cyber Extortion" events are more severe in capital intensive companies (companies with high revenue and low number of employees). Data physically stolen and unintentionally disclosed are less severe in the USA, what might be a result of the more restrictive data protection policies in the USA in comparison to the rest of the world. The financial and healthcare sectors seem to suffer less severe losses in almost all cyber risk categories, except for ``Data - Unintentional Disclosure", where the financial sector seems to be more exposed than other business sectors. The last two columns of Table~\ref{table:model0_sigma} show the impact of contagion on cyber event severity. Data malicious breaches, caused by or causing other cyber events in the same company, have the potential to trigger heavier financial losses. While for the risk types ``Privacy - Unauthorized Contact or Disclosure" and ``Identity – Fraudulent Use/Account Access" spillover events in the same company are generally less severe. This may indicate that for these particular risk categories companies have cyber risk management practices in place that reduce losses from repeated breaches. Events linked to external enterprises increases the severity for the categories ``Privacy – Unauthorized Data Collection" and ``Data - Physically Lost or Stolen". 
\begin{table}[h]
    	\centering
    	\caption{This table reports the regression score ratios for parameters in the link function $\log(\mu)$. The significance of each coefficient is reported with the usual convention  of 1, 2, and 3 stars corresponding to a p-value of $10\%$, $5\%$, and $1\%$, respectively.}
    	\label{table:model0_mu}
        \begin{adjustbox}{center,scale=0.55}
        	\begin{tabular}{llllllllllll}
        		RiskType & $\beta_0$ & Time        & $R_{medium}$      &	$R_{big}$    &	$E_{medium}$   &	$E_{big}$  &	$L_{USA}$	      & $B_{financial}$       & $B_{health}$        & ML       & MC       \\
        		
        		\hline

                    Privacy - Unauthorized Contact or Disclosure          &6.3037***	 & -6.2111*** & -1.6977*  &	2.6485** & 0.3417   & 1.9169**   &	3.859***   &	8.413***   & -0.0879    &	           & -4.1663***\\
                    Privacy – Unauthorized Data Collection              &4.2443***	 & -4.2744*** & 5.7174*** &7.9349*** &	-4.256  & -8.1955*** &	3.4149***  &	13.7316*** & 4.4548***  &   12.3974***   &	     \\
                    Data - Physically Lost or Stolen         &1.1337	 & -1.1599    & -0.5047   &-0.7503   &	5.1968  & 5.2282***  &	-1.5124*   &	5.5719***  & 3.9608***  &		       & -1.591 \\
                    Identity – Fraudulent Use/Account Access               &1.2702	 & -1.2843    & 2.7723*** &-2.6023*  &	0.6804  & 3.8916***  &	-3.083***  &	-2.8866*** & -1.4372    &	1.0646     &	     \\
                    Data - Malicious Breach                  &-7.1916*** &	7.3144*** & 0.4154	  &3.099***	 &-1.1811   & 0.1218     &	5.3967***  &	3.0396***  & 4.5634***  &	-6.6742*** & -11.0582***\\
                    Phishing, Spoofing, Social Engineering                   &-6.3192*** &	6.3203*** & 4.6131*** &0.1463    &	-0.9365 & -1.5422*   &	-1.8377**  &	1.2422     & 7.8563***  &		       &        \\
                    IT – Configuration/Implementation Errors &6.0355***	 & -6.0597*** & 7.821***  &3.6946*** &	3.6344  & -4.1544*** &	-0.2476    &	3.4372***  & 5.2348***  &	6.6952***  &         \\ 
                    Data - Unintentional Disclosure          &4.354***	 & -4.3879*** & 0.4246	  &-1.6519** &	3.3095  & 3.9499***  &	-1.404*    &	-2.3172**  & 4.3793***  &		       &       \\
                    Cyber Extortion                          &-16.203*** &	16.1703***& -4.9287***& 0.8184   &	2.1502  & -0.9929    &	0.867      &	3.7901***  & 10.8866*** &		       &           \\
                    Network/Website Disruption               &-4.3018*** &	4.2737*** & 2.5043**  &1.7096*   &	-2.2092 & -0.6527    &	0.6316     &	0.3401     & 6.0102***  &		       &           \\
                    Skimming, Physical Tampering          &1.7656*	 & -1.7583*   & -2.642**  &-1.0112   &	-0.5355 & 0.9971     &	-5.8438*** &	-1.5343*   & 3.4987***  &	-4.201***  &           \\
                    IT - Processing Errors&-1.0354           &	1.0261    & 0.4021	  &0.3733     &	0.6079  & -1.3304    &	4.3801***&	3.2282***  &               &            &                         \\		
                    Industrial Controls and Operation        &-1.9461*   &	1.9537*   &           &          & -2.8584  &            &	           &		       &	        &              &            \\
                    Undetermined/Other                       &0.8327	 & -0.8327    & 1.8804    &	2.0735** &	-1.6645 & -1.8697*   &	1.1559     &	1.479*     &            &	           &           \\

        		\hline	

        	\end{tabular}
 	    \end{adjustbox}
    \end{table}
\begin{table}[h]
    	\centering
    	\caption{This table reports the regression score ratios for parameters in the link function $\log(\tau)$. The significance of each coefficient is reported with the usual convention  of 1,2, and 3 stars corresponding to a p-value of $10\%$, $5\%$, and $1\%$, respectively.}
    	\label{table:model0_sigma}
    	\begin{adjustbox}{center,scale=0.55}
        	\begin{tabular}{llllllllllll}
        		RiskType & $\beta_0$ & Time        & $R_{medium}$      &	$R_{big}$    &	$E_{medium}$   &	$E_{big}$  &	$L_{USA}$	      & $B_{financial}$       & $B_{health}$        & ML       & MC       \\
        		
        		\hline
                    Privacy - Unauthorized Contact or Disclosure 	         &6.6067*** &	-6.5183*** &	-0.1413   &	3.0814*** &	-1.0599    &	0.0451     &	-0.5747   &	6.635*** &	1.7084**  &		     & -2.6275***\\
                    Privacy – Unauthorized Data Collection             &4.3034*** &	-4.3115*** &	6.0298*** &	4.7691*** &	-4.3361*** &	-6.3732*** &	0.1952	  & 9.8741***&	5.0142*** &	8.1338***&	        \\
                    Data - Physically Lost or Stolen         &0.8894    &	-0.8623    &	-0.2714   &	-0.8924   &	2.6821***  &	3.139***   &	-2.2156** &	1.6157** &	2.1862**  &  		 &-3.2789*** \\
                    Identity – Fraudulent Use/Account Access               &2.7085*** &	-2.7197*** &	1.3889    &  -1.0612  &	 1.2192    &	1.2382     &	0.9968    &	-1.273   &	-1.7623*  &	2.9839***&	       \\
                    Data - Malicious Breach                  &-2.3488** &	2.4646**   &	-0.5338   &	-0.1638	  &-0.0974     &	-0.3404    &	2.3161    &	1.0256	 &  7.1538*** &	-4.072***&	-5.1573\\
                    Phishing, Spoofing, Social Engineering                     &-5.2824*** &	5.2772***  &	4.9515    &	-1.6083	  &0.9595      &	-0.8269	   &    1.6028	  & 2.1316** &  7.6076*** &		     &\\
                    IT – Configuration/Implementation Errors &2.7334*** &	-2.7344*** &	7.8107*** &	1.3124    &	6.3982***  &	-3.8609*** &   -0.8266    &	0.4187   &  3.4246*** &	3.3014***&	\\
                    Data - Unintentional Disclosure          &3.3412*** &	-3.3319*** &	1.0442    &	-1.5094	  &1.9885**    & 	1.572      &	-7.3631***&	-1.6987**&	3.7304*** &		     &\\
                    Cyber Extortion                          &-5.3688***& 	5.3695***  &	-3.6261***&	-2.5504** &	2.8859***  &	-0.9208	   &     0.2467***&	1.797*** &  12.2652***&		     & \\
                    Network/Website Disruption               &0.6384    &	-0.657     &	3.2336*** &	-0.8623   &	-1.2996	   &     -1.3532   &	2.0436**  &	0.5777   &	4.2778*** &		     &\\
                    Skimming, Physical Tampering          &2.9529*** &	-2.9405*** &	-2.3306** &	-0.0658   &	1.0398	   &    -0.971     &	-3.2267***&	0.6312   &	8.7332*** &	-1.0688  &	\\
                    IT - Processing Errors                   &-1.683*   &	1.6756*    &	3.8958*** &	3.1338*** &  -0.3692   &	-3.5478*** &	2.2917**  & 1.0939	 &		      &          &    \\ 
                    Industrial Controls and Operation        &0.0042    &	-0.0004    &              &	          &	0.0003     &               &              &          &            &		     &		\\		
                    Undetermined/Other                       &-0.1794   &	0.1891	   &    -0.0265   &	-0.024    &	0.1261     &	-0.0299    &	-0.0431	  &-0.0355	 &            &	         & 	\\

        		\hline	
        
        	\end{tabular}
    	\end{adjustbox}
    
    \end{table}
    
Additional analysis on the interaction effect among dummy covariates and Q-Q plot can be found in Section~\ref{appendix:non_linearity_interaction} of the Appendix. 
\clearpage
\section{Rank-based Regression of Cyber Risk Loss Processes}
\label{Sec:ranked_based_regression}
Results from both the non-parametric tail estimation and the GAMLSS regression analysis suggest that cyber event severity exhibits extreme heavy tails, with no finite first moments and no easily identifiable structure. Given the data structure, it is also possible that our results for the identified drivers of cyber risk are at least partially driven by few very extreme observations in the individual risk categories. We therefore seek to extend our analysis by applying an additional rank-based approach that removes the scale of the effect of extreme loss events. Thus, we apply an ordinal regression framework in order to control for the leverage effects that may be present in the regression estimators that could arise from the extreme magnitude of cyber risk losses. Ordinal regression can also be seen as an additional robustness check on the statistical significance of the covariates in the combined POT and GAMLSS approach.

Ordinal regression analysis have been implemented in the context of cyber risk in various settings, where often data is expressed in terms of ordered level of severity \cite[see, among others][]{raffinetti2015,giudici2020cyber,sexton2015,hubbard2010}. Typically, cyber event severity $Y$ can be transformed into its rank $R$, and then explained by a linear regression model. By construction, the rank transformed response variable does not present heavy tails anymore. Let $Y$ to be expressed in $k$-levels, then cyber severity can be transformed into ranks according to the following steps: $r_1 = 1$ corresponds to the rank of the smallest category, $r_z = n_{z-1}+r_{z-1}$, with $n_z$ being the absolute frequency of rank $r_z$, and $z=1,\dots,k$; see \cite{iman1979}. Then the impact of covariates on the rank transformed cyber event severity can be addressed with the following linear regression:
\begin{equation}
\label{eq:rank_regression}
    R=  \beta_0+\sum_{s=1}^{S} \beta_s RT_s+\beta_tt+\sum_{l\in \mathcal{L}} \beta_lL_l+\sum_{b\in \mathcal{B}}\beta_bB_b+\sum_{r\in\mathcal{R}} \beta_rR_r+\sum_{e\in \mathcal{E}}\beta_eE_e+\beta_{ml}ML+\beta_{mc}MC+\epsilon,
\end{equation}
with $\mathbb{E}[\epsilon]=0$ and $\mathrm{var}(\epsilon)=\sigma^2$. To evaluate the goodness of fit of the regression in~(\ref{eq:rank_regression}), we employ the cross-validation method based on the Rank Graduation Accuracy \cite[see,][]{giudici2020cyber}:
\begin{equation*}
    RGA_{R} = \sum_{i=1}^n\frac{n}{i}\left(\frac{1}{n\Bar{r}} \sum_{j=1}  r_{ord(\hat{r}_j)}-\frac{i}{n}\right)^2, 
\end{equation*}
where $\Bar{r}$ is the average rank, $r_{ord(r_j)}$ is the rank transformed response variable reordered according to the predicted rank $\hat{r}_j$. $RGA_{R}$ is based on the concordance curve given by the pairs $\left(i/n,\frac{\sum_{j=1}^iy_{r_j}}{n\Bar{y}}\right)$, for $i =1,\dots,n$. One interprets $RGA_R$ as follows, a large value implies a better fit with respect to a random model with no predictive power and a lower value implies the presence of a model that may not be readily distinguished from an arbitrary random model with no predictive power.

\begin{figure}[ht]
	\centering
	\subfloat[{Concordance curve of rank-transformed response variable }]{{\includegraphics[width=0.45\textwidth,height=0.3\columnwidth]{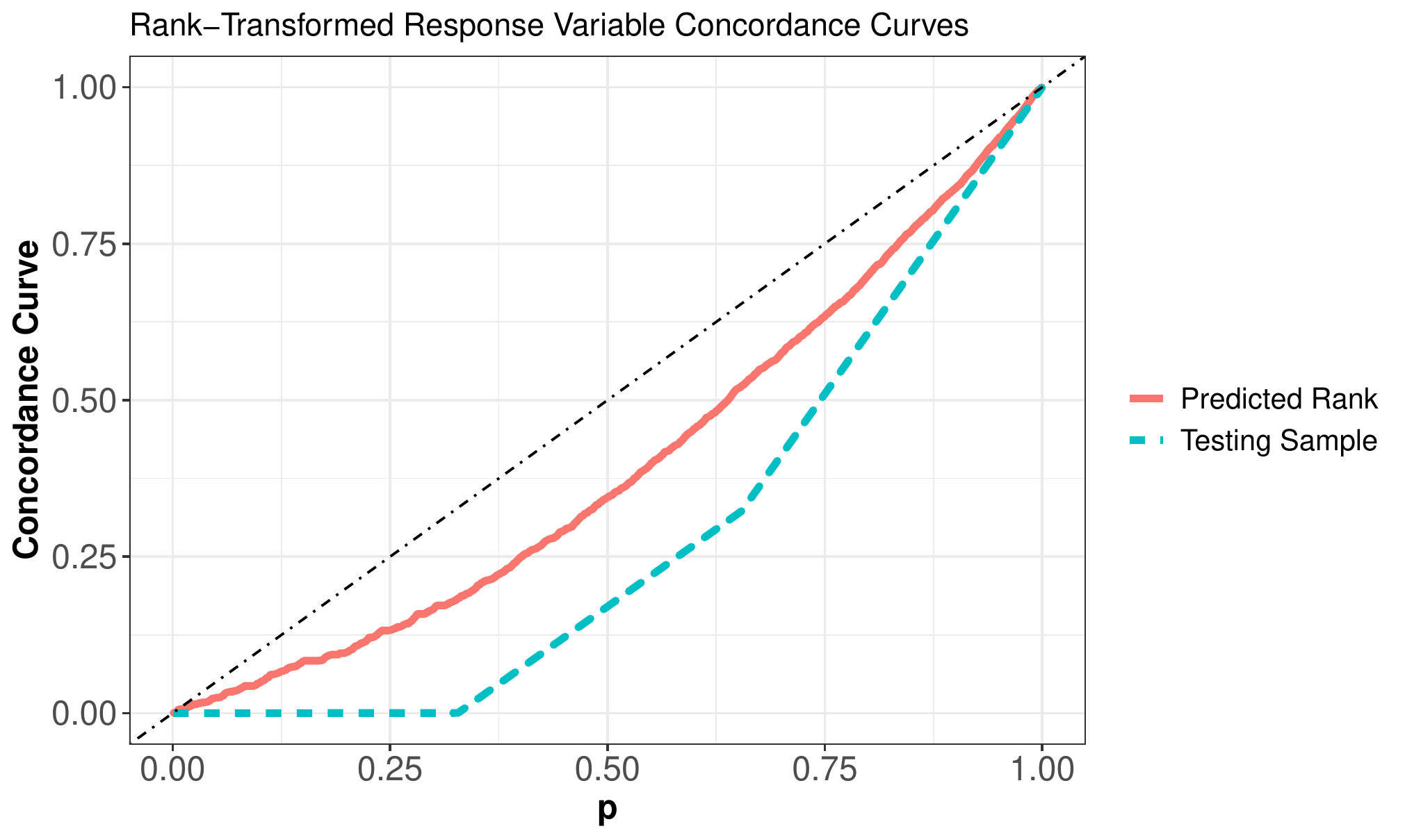}}}
	\qquad
	\subfloat[{Concordance curve of response variable}]{{\includegraphics[width=0.45\textwidth,height=0.3\columnwidth]{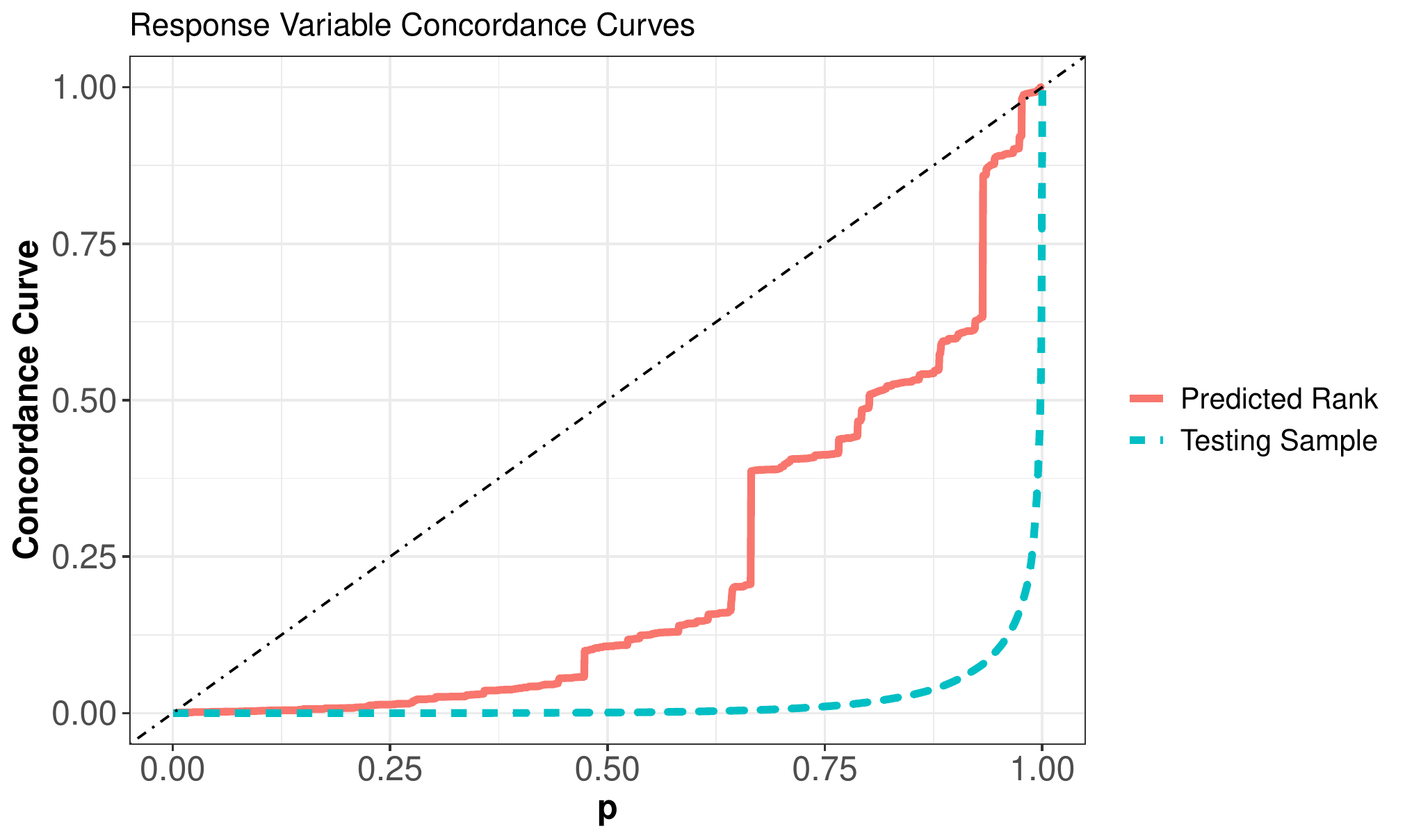}}}
	\caption[Concordance Curve]{This figure shows the concordance curves of cyber event severity, in terms of rank-transformed response variable (panel (a)) and response variable (panel (b)). Rank-transformed response variable and response variable are reordered according to the predicted rank in the training sample. }
	\label{Fig:concordance_curve}
\end{figure}

Figure~\ref{Fig:concordance_curve} shows the concordance curve in terms of rank-transformed response variable (panel (a)) and  in terms of the response variable reorder according to $\widehat{r}_j$ (panel (b)) for the linear model in (\ref{eq:rank_regression}). The bisector corresponds the the concordance curve of a model with no explanatory power (random model), where all the observation share the same predicted rank.  In other words, the further away a concordance curve is from the bisector, the better the performance of the corresponding linear model. 
The rank transformed concordance curves (in red in both panel (a) and (b)) lies below the bisector, implying the linear model in (\ref{eq:rank_regression}) has a better predictive power than a random model.

To test for the significance of the regression covariates in the ordinal regression we adopt the  testing methodology based on the $RGA_R$ in~\cite{giudici2020cyber, raffinetti2015}. Here, we report only the main findings, a detailed description of the procedure and results can be found in Section~\ref{appendix:rga_test} of the Appendix. We first proceed in testing the significance of each coefficient of the linear model in (\ref{eq:rank_regression}). The outcome of this analysis indicates that that time,  medium number of employees and USA are the only statistical significant dummy covariates. Furthermore, we found that cyber risk types are not statistical significant, neither jointly nor separately. In terms of RGA, all the tested models present similar values of RGA, with the model where the covariate controlling for high revenue companies ($R_{big}$) is omitted scoring the highest. These findings seem to suggest that the results on the significance of covariates in the previous section involving the study of the GAMLSS regression are mainly driven by extreme cyber events, an outcome we suspected could be the case and which led us to explore the ordinal regression as a means of confirmation.

Finally, as a robustness check on the result of the Voung's variance test, we test whether the rank transformed response variable should be fitted jointly on the risk categories, or separately. In other words, we perform the following test of hypothesis:
\begin{equation}
    H_0: \text{Joint Model}\qquad \text{vs}\qquad H_1:\text{Separate Model}\notag
\end{equation}
where in the separate model, the 
 rank-based regression is as follows:
\begin{equation}
    R_s=  \beta_0+\beta_tt+\sum_{l\in \mathcal{L}} \beta_lL_l+\sum_{b\in \mathcal{B}}\beta_bB_b+\sum_{r\in\mathcal{R}} \beta_rR_r+\sum_{e\in \mathcal{E}}\beta_eE_e+\beta_{ml}ML+\beta_{mc}MC+\epsilon,\notag
\end{equation}
where $R_s$ is the rank transformed response variable of risk type $s$. We find a significance value above 80\%, implying that the models are almost always significantly different. This confirms once again the evidence to further support our conjecture regarding the findings in the previous sections, that one must be careful regarding the indifference one can take to the joint versus decoupled approaches due to the sensitivity that this analysis may have to the extreme losses in the sample. In Section \ref{SubSec:JointDecoupled} it was determined that cyber event severity is so heavy tailed that it's not possible to distinguish between a joint and a separate model. However, once the effects of extreme events are removed through a rank based ordinal regression analysis, the conclusion changes and it is no longer the case that one can say that the joint and decoupled modelling approaches are indistinguishable. Moreover, it implies that the Voung's variance test result is more likely to be driven by extreme events in the loss data, rather than the independence assumption made in Section~\ref{sec:model}. Consequently, we have included further studies on the joint model for completeness of the analysis in Sections~\ref{appendix:rga_test} and~\ref{appendix:gpd_vs_lognormal} of the Appendix. 

As a conclusion to this debate, it comes down to what types of data one wishes to make inference about. If the raw monetary loss amounts are under consideration then this debate regarding joint versus decoupled is shown to be inconsequential in the aforementioned analysis. However,
in the case where the relative severity effects are under investigation, which might be useful for management purposes and comparative analysis
of effectiveness of certain management and business decisions relating to cyber mitigation for example, one may resort to model ranks. Then in this case, one will need to consider the joint modelling framework and decoupled framework as distinguishable and then perform analysis of each case.

Hence, in analyzing cyber event severity by the means of monetary losses, one can either proceed with the joint or the decoupled approach, the latter having the advantage of superior flexibility that can capture cyber risk heterogeneity. When instead one wishes to model rank transformed cyber severity, either for filtering out the effect of extreme events or due to other specific data characteristics, one needs to take a decision on which approach is more statistically sounded. Such decision can then be made using the $RGA_R$ and concordance curve framework, or any other sensible accuracy measure.

\section{Addressing Cyber Risk Capital and Insurability via GAMLSS Regression Models}
\label{Sec:case_study}
In this section we present two case studies that will utilise the GAMLSS regression models in the context of both risk management and insurance. The first case study is a Value-at-Risk calculation based on the estimates obtained in Section~\ref{Sec_gamlss_analysis}. The second application is related to insurance premium calculations, based on Advisen data and our estimates.

\subsection{Value-at-Risk for Cyber Risk Capital Reserves under GAMLSS Regression Model}
\begin{figure}[ht]
	\centering
	\subfloat[{Fitted values of $\log(\widehat{\lambda})$ }]{{\includegraphics[width=0.45\textwidth,height=0.3\columnwidth]{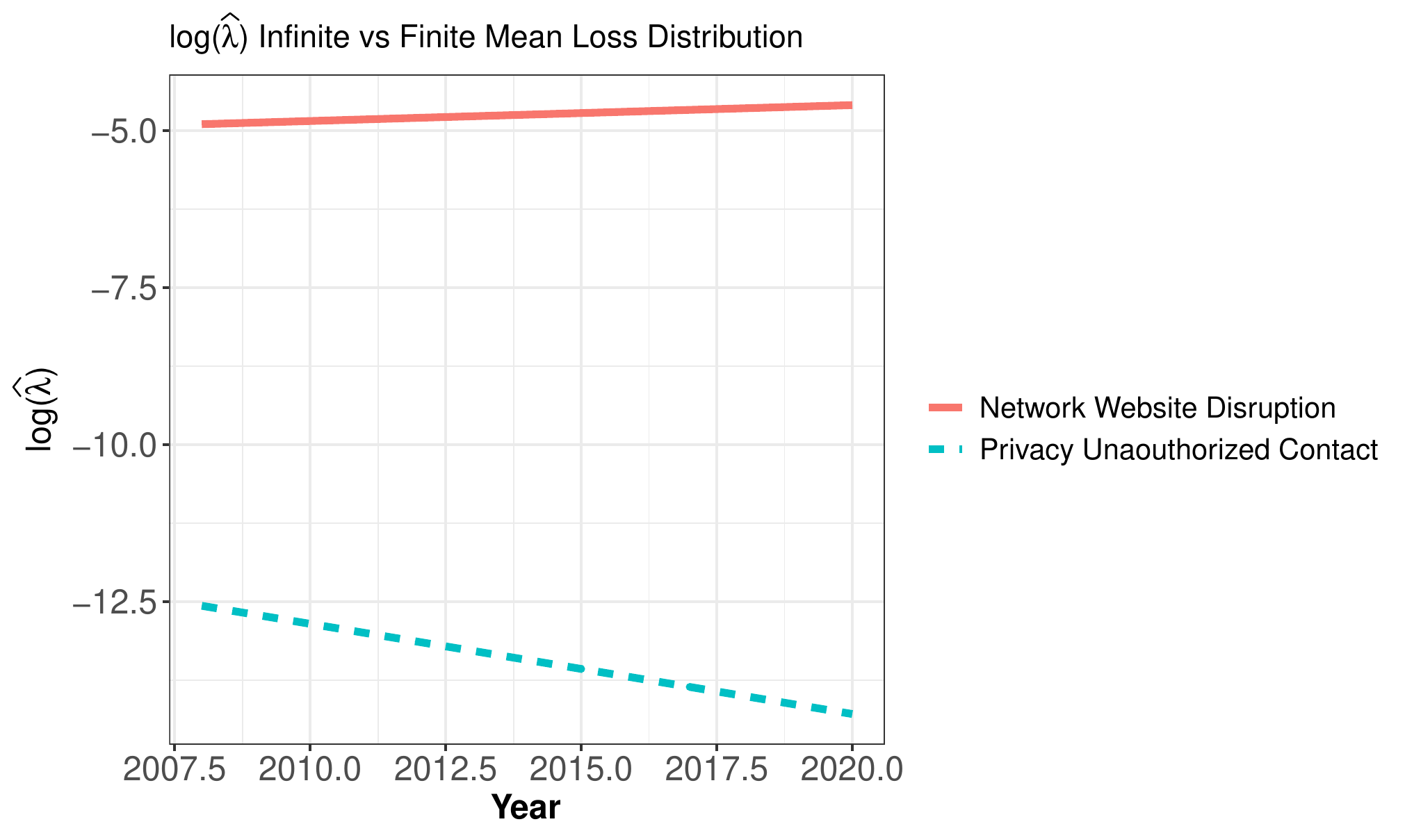}}}
	\qquad
	\subfloat[{Fitted Values of $\log(\widehat{\tau})$}]{{\includegraphics[width=0.45\textwidth,height=0.3\columnwidth]{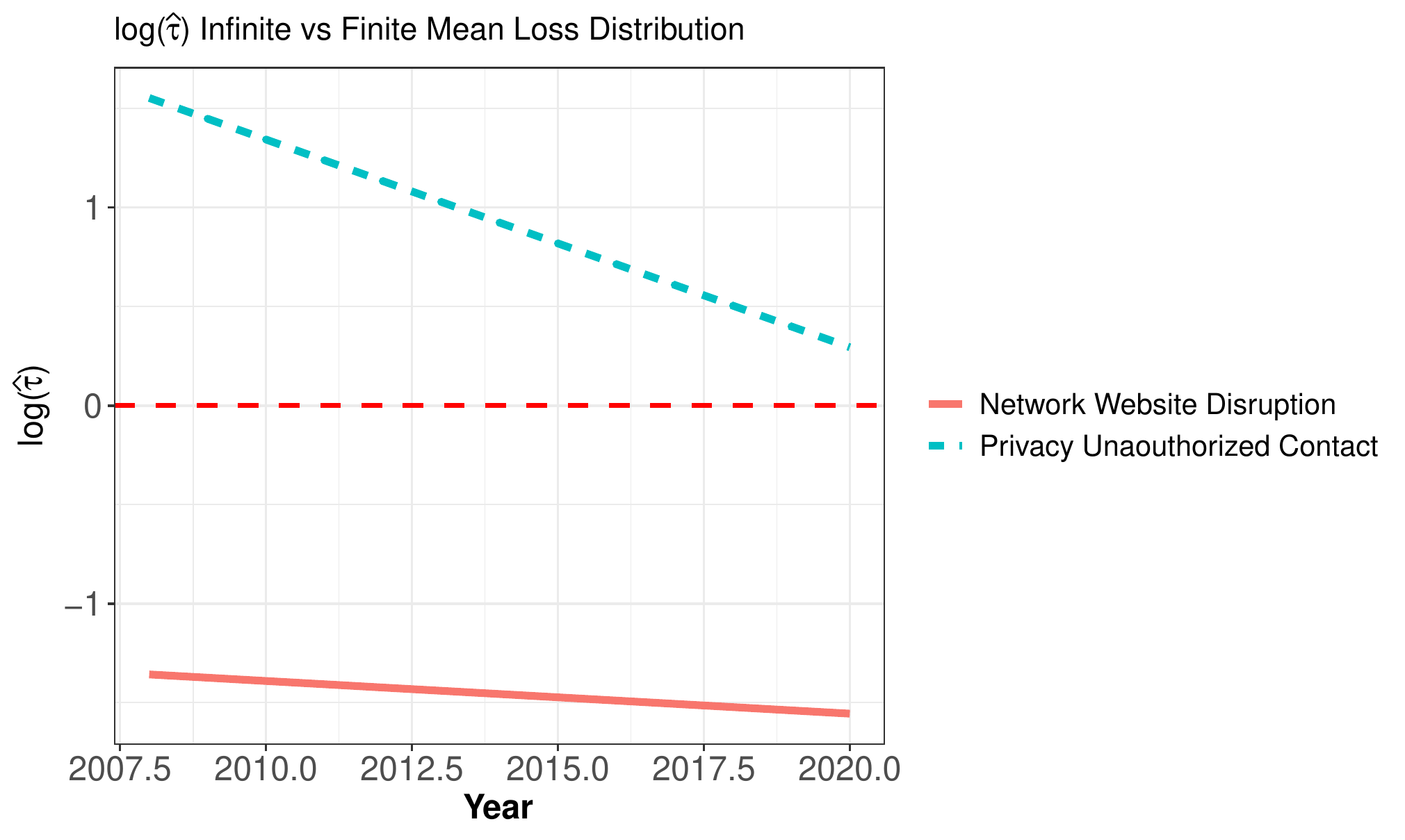}}}
	\qquad
	\subfloat[{Value-at-Risk }]{{\includegraphics[width=0.45\textwidth,height=0.3\columnwidth]{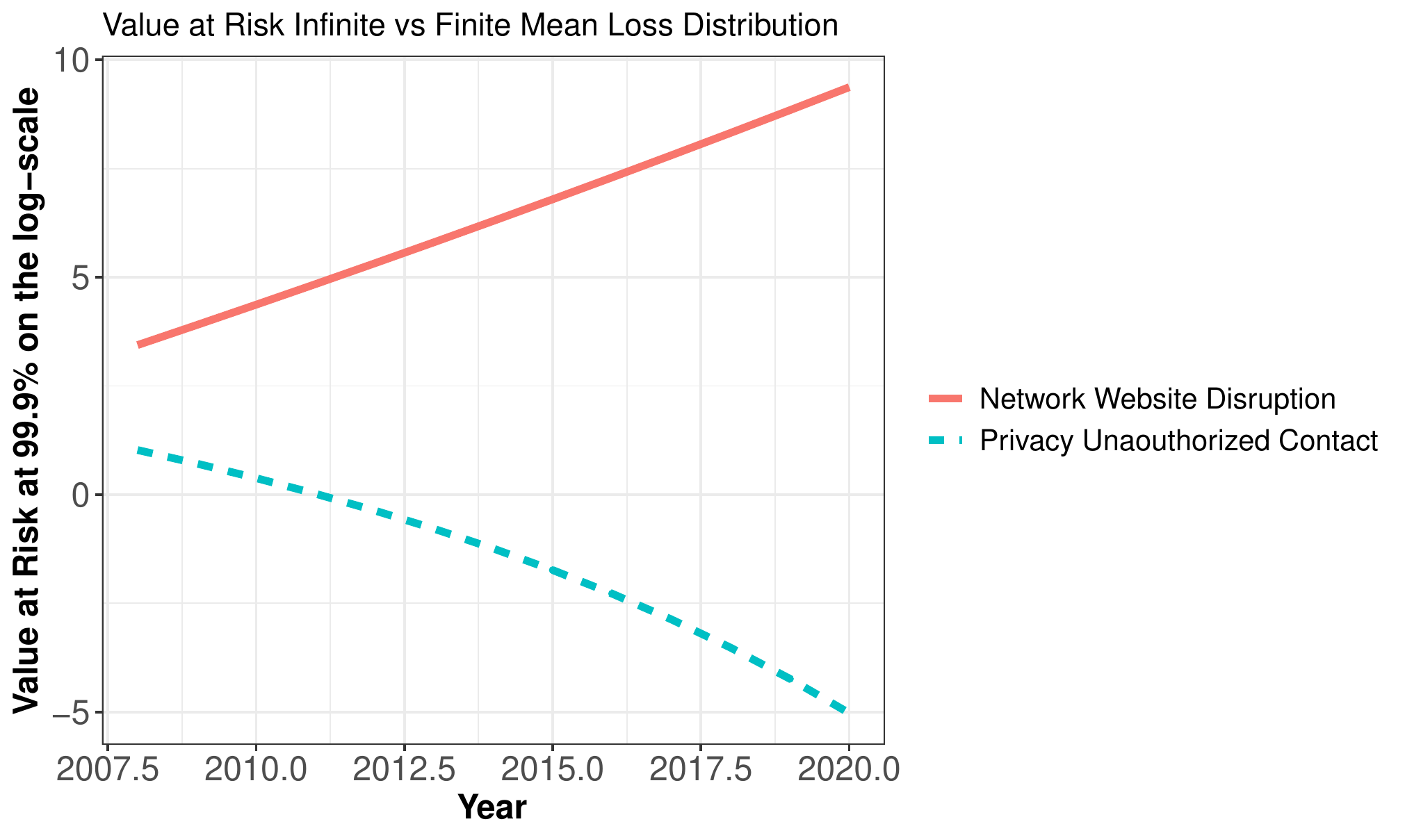}}}
	\caption[VaR Case 1]{This figure shows the  Value-at-Risk and fitted values for $\lambda,\tau$ of ``Network/Website Disruption" in red and ``Privacy - Unauthorized Contact or Disclosure" in blue for the hypothetical company. The values of $\tau$ under the red dashed line in panel (b) correspond to the infinite mean case. Capital requirements for ``Network/Website Disruption" are much higher in the case of ``Privacy - Unauthorized Contact or Disclosure".}
	\label{Fig:VaR_case1}
\end{figure}

The three sets of analysis undertaken in previous sections have provided strong statistical evidence to assert that cyber risk related losses are heavy tailed. It is well known that heavy tails have serious implications for capital requirement calculations \cite[see, e.g.][]{nevslehova2006}. Estimates in Tables~\ref{table:model0_lambda}, \ref{table:model0_mu} and \ref{table:model0_sigma}  show that cyber event frequency and severity depend on cyber threat types and company characteristic. To illustrate the impact of covariates on capital requirement we present the Value-at-Risk calculation using the Single Loss Approximation (SLA); see \cite{degen2010}, \cite{peters2013understanding} and book length reviews in \cite{peters2015advances, cruz2015fundamental}:
\begin{equation}
\label{eq:Var}
    \mathrm{VaR}_\alpha (Z) \approx u + GPD^{-1}\left(1-\frac{1-\alpha}{\widehat{\lambda}};\widehat{\mu},\widehat{\tau}\right) \left(1-\dfrac{(1-\alpha)c(\widehat{\tau})}{1-\widehat{\tau}}\right), 
\end{equation}
where $c(\widehat{\tau}) =\frac{1}{2}(1-\widehat{\tau})\frac{\Gamma^2(1-\widehat{\tau})}{\Gamma(1-2\widehat{\tau})}$, $GPD^{-1}\left(\cdot;\widehat{\mu},\widehat{\tau}\right)$ is the inverse  distribution function of a GPD random variable with parameters $\widehat{\mu}$ and $\widehat{\tau}$, and $\Gamma(\cdot)$ is the gamma function. Using the fitted values of $\lambda,\mu$ and $\tau$ from the GAMLSS approach in (\ref{eq:Var}), it is possible to relate specific business models and structures to capital requirement for any give cyber risk type. In line with the discussion of the previous section, we present two hypothetical examples to illustrate how the GAMLSS based SLA for this regression based LDA model is instructive for risk management and capital reserving. The first example shows the impact on capital requirement of a fitted value of $\tau$ lower than 1. The second one focuses on the effects of contagion, allowing cyber related events of the type ``Data - Malicious Breach" to be linked to multiple losses in various companies. Figure~\ref{Fig:VaR_case1} depicts the fitted values of $\lambda$, $\tau$, and the corresponding Value-at-Risk of two risk types (``Network/Website Disruption" in red and ``Privacy - Unauthorized Contact or Disclosure" in blue) for one company, residing outside the USA, operating neither in financial nor in healthcare sector, and having high revenue and high number of employees.

Panel (a) of Figure~\ref{Fig:VaR_case1} shows the fitted values of $\lambda$ on the log scale. ``Network/Website Disruption" frequency was higher than ``Privacy - Unauthorized Contact or Disclosure" frequency during the period 2008-2020. Moreover,  ``Network/Website Disruption" frequency is following an increasing trend, while ``Privacy - Unauthorized Contact or Disclosure" frequency is decreasing over time, in agreement with the score ratios in Tables~\ref{table:model0_lambda}. Looking at panel (b), both the risk types show decreasing trends in the values of $\log(\widehat{\tau})$, meaning that their severity distributions have become more heavy tailed over time, with an increasingly higher probability of extreme events. What really sets the difference between these two risk types is that ``Network/Website Disruption" has $\log(\widehat{\tau})$ lower than 0, implying that the corresponding distribution has no finite first moment, and therefore, events from such type can generate infinite expected loss. This last aspect is further shown on panel (c) with the Value-at-Risk (on the log scale) computed using the formula (\ref{eq:Var}). The Value-at-Risk for ``Network/Website Disruption" is increasing over time following what it seems to be an exponential growth path, and indicating that the capital required in the case of infinite mean loss distribution is much higher than in the case of ``Privacy - Unauthorized Contact or Disclosure".

\begin{figure}[ht]
	\centering
	\subfloat[{Fitted values of $\log(\widehat{\lambda})$ }]{{\includegraphics[width=0.45\textwidth,height=0.3\columnwidth]{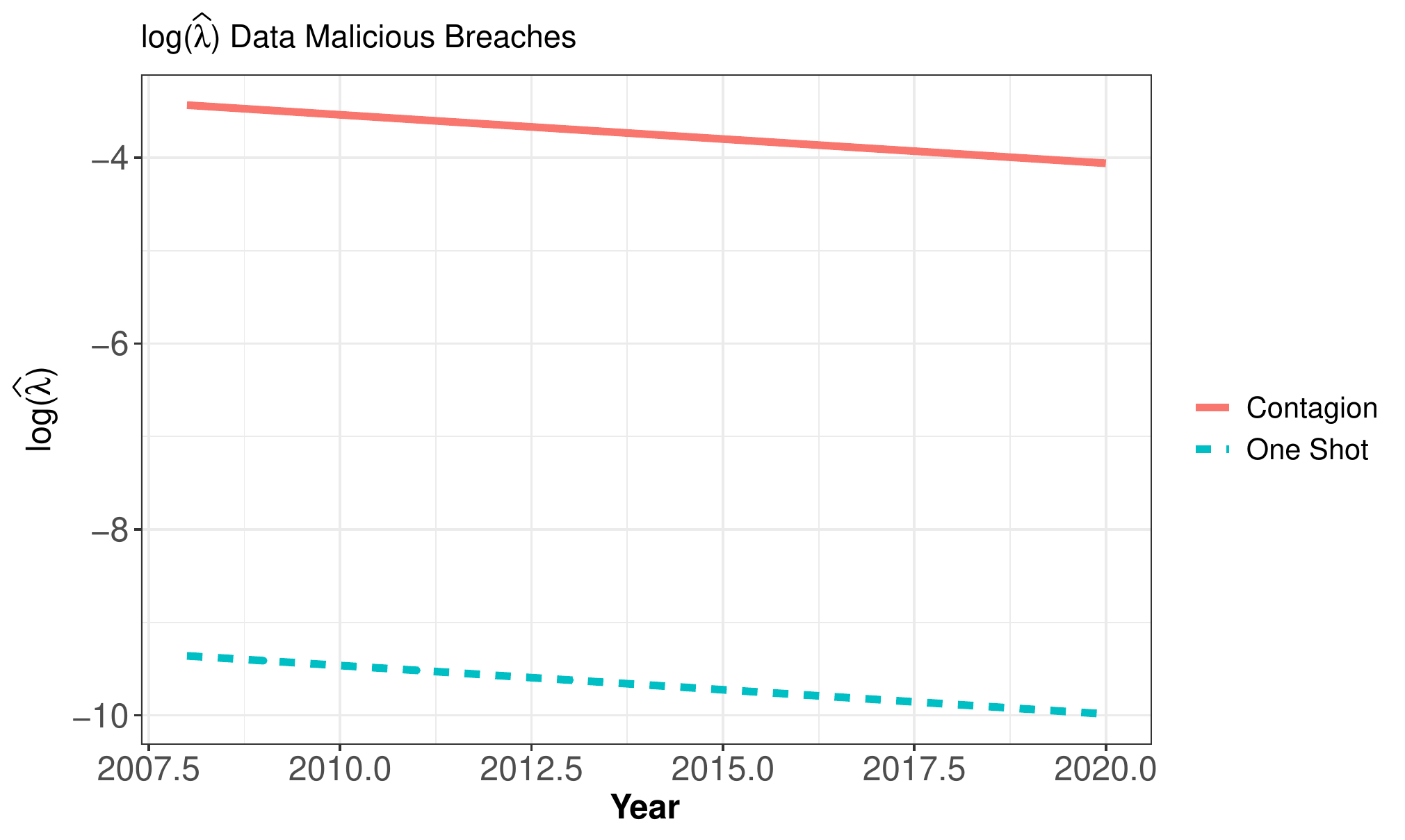}}}
	\qquad
	\subfloat[{Fitted Values of $\log(\widehat{\tau})$}]{{\includegraphics[width=0.45\textwidth,height=0.3\columnwidth]{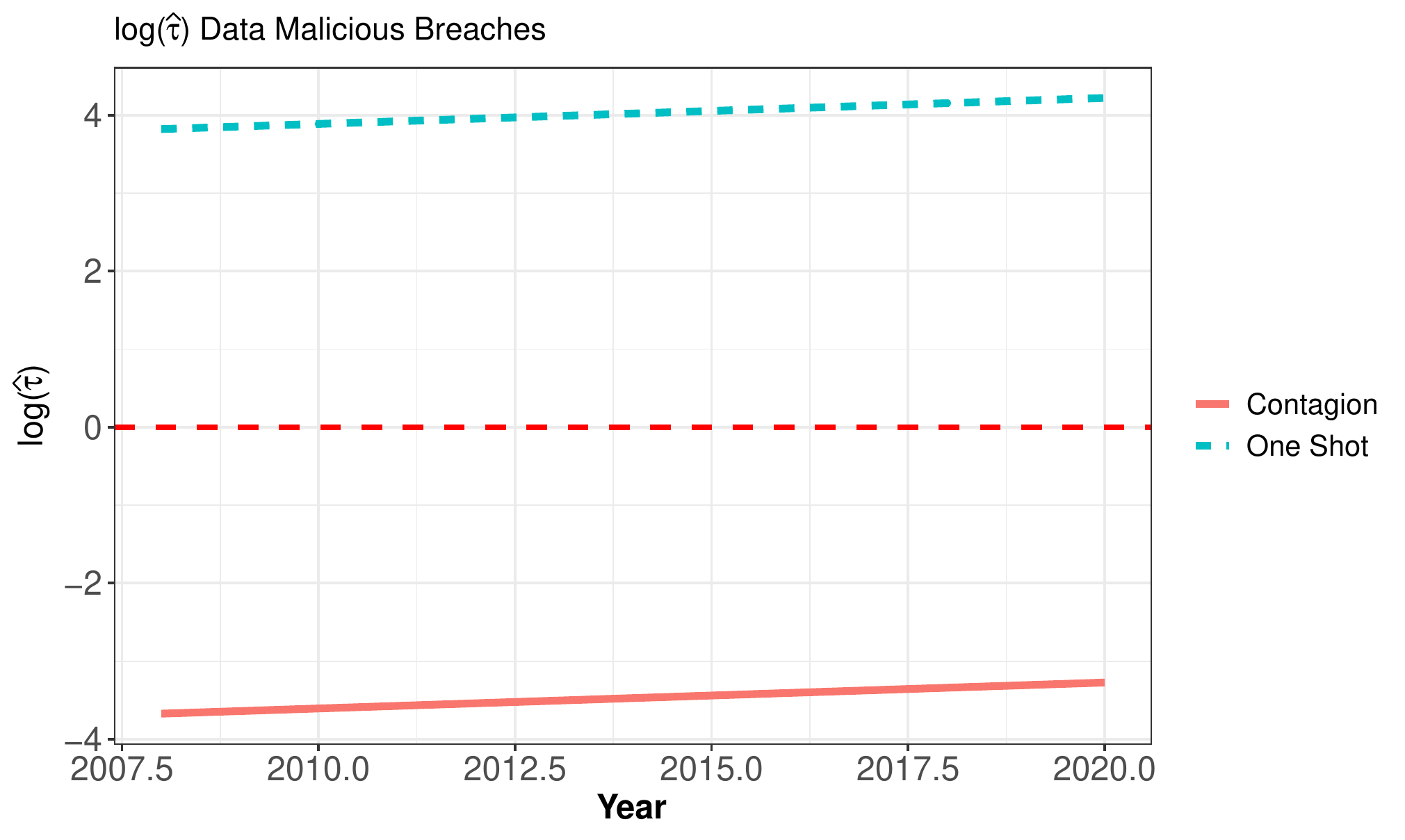}}}
	\qquad
	\subfloat[{Value-at-Risk }]{{\includegraphics[width=0.45\textwidth,height=0.3\columnwidth]{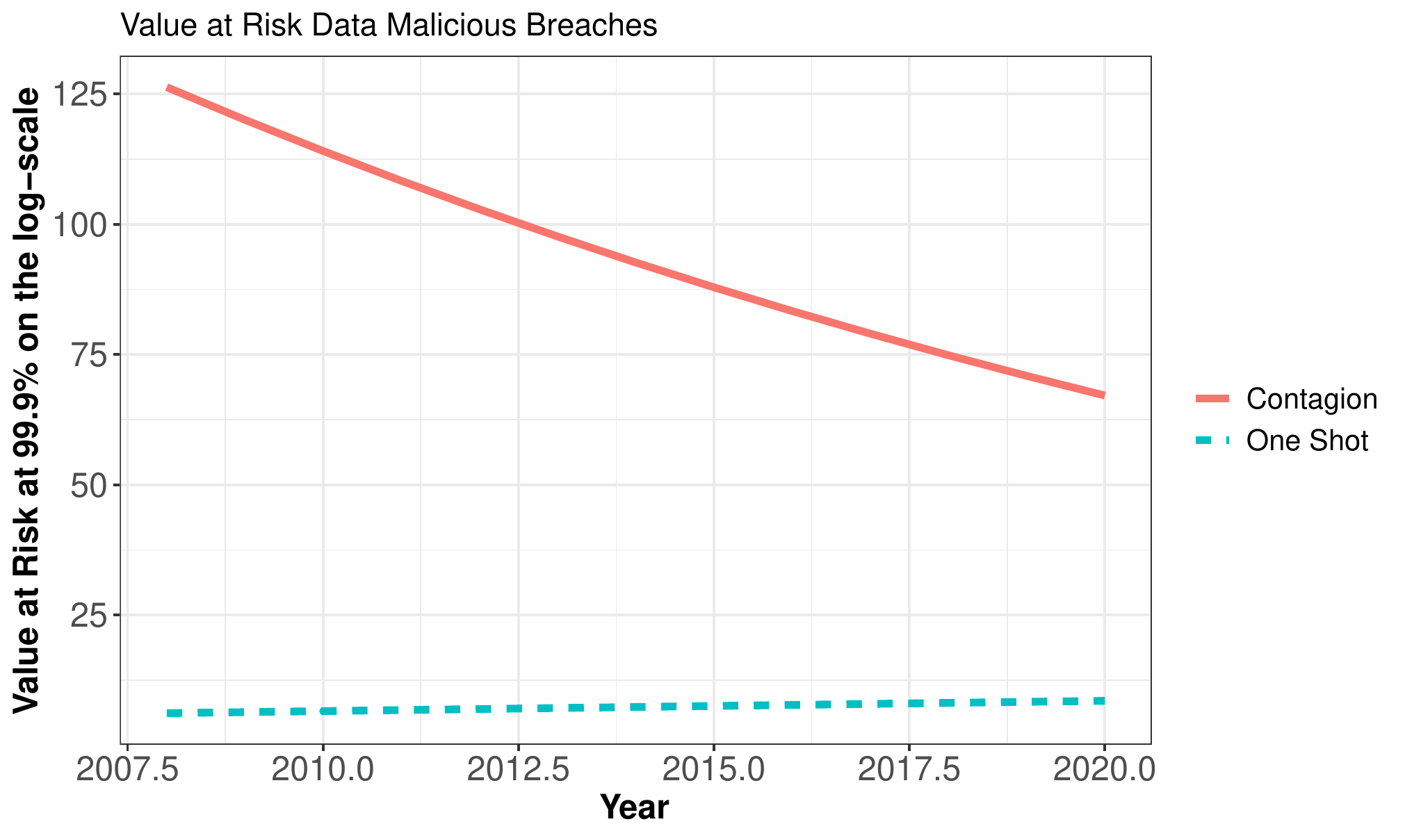}}}
	\caption[VaR Case 1]{This figure shows the  Value-at-Risk and fitted values for $\lambda,\tau$ for ``Data - Malicious Breach". The values of $\tau$ under the red dashed line in panel (b) correspond to the infinite mean case. Capital requirements for ``Data - Malicious Breach" due to contagion type of events are much higher than those for one shot events. }
	\label{Fig:VaR_case2}
\end{figure}

The effect of contagion on ``Data - Malicious Breach" is shown in Figure~\ref{Fig:VaR_case2}. Figure~\ref{Fig:VaR_case2} shows the fitted values of $\lambda$, $\tau$, and the corresponding Value-at-Risk for two companies, both residing outside the USA, having medium revenue and high number of employees, and suffering from multiple events in red (henceforth company A) and suffering only from one shot events in blue (henceforth company B). 
Data malicious breaches are more frequent and severe in company A than B, having higher fitted values of $\lambda$ and lower fitted values of $\tau$. Considering contagion type of events, increases the riskyness of data malicious breaches considerably, since the corresponding values of $\log(\widehat{\tau})$ are shifted downwards below the threshold level of one, turning a cyber risk types with finite first moments into the infinite mean loss distribution case.
These two hypothetical examples shows the importance of considering covariates into capital requirement calculation. The heterogeneous nature of cyber risk is then reflected into the variations in capital requirements for different cyber risk types. Similar results  can be obtained with other combination of covariates and risk types.

\clearpage

\subsection{Cyber Risk Insurance}
The statistical features of cyber risk have important implication on modelling and on prudential capital requirements. In this section we address the implication of our findings on the insurabiltiy of cyber risk. We show using a simple insurance premium calculation that cyber risk reveals to be a formidable foe from the insurance perspective, having many aspects that undermine its insurability.

Consider a non satiable risk averse decision maker with total wealth $w$, facing the possibility to suffer from  $\widetilde{Y_1},\dots,\widetilde{Y_N}$ random losses, where $N$ follows a poisson distribution. The decision maker is offered 1 year insurance policy to protect against each random loss up to a top cover limit equivalent to a percentage of the company wealth. According to the zero utility principle, the maximum premium he is willing to pay will be the solution $P^+$:
\begin{equation}
\label{eq:decision_premium}
    \mathbb{E}\left[u\left(w-\sum_{i=1}^N\widetilde{Y}
    _i\right) \right] = \mathbb{E}\left[u\left( w-P^+ -\sum_{i=1}^N\widetilde{Y}_i+\sum_{i=1}^N\min(\widetilde{Y_i},kw)\right)\right],
\end{equation}
where $u(x)=\widetilde{u}\left(\max(x,1)\right)$, $\widetilde{u}$ is a concave, non decreasing utility function, and $k$ is the percentage corresponding to the top cover limit. The minimum premium an insurer with utility function $v$ and wealth $W$ is willing to accept to insure the decision maker is given by $P^-$ solution of the following non linear equation:
\begin{equation*}
    v(W) = \mathbb{E}\left[v\left(W +P^--\sum_{i=1}^N\min(\widetilde{Y_i},kw)\right)\right]. 
\end{equation*}
The decision then to insure and to be insured depends, among other things on the decision maker and insurance company total wealth. To get rid of one dimension, we consider the hypothetical case where  $P^+$ is the equilibrium premium (i.e. $P^+=P^-$) and then we focus on \emph{how big} the insurance company should be in order to insure the decision maker. In other words, assume that  $W=mP^+$, where $m$ is the size of the insurance pool. Then, the problem for the insurer becomes, finding the optimal pool size, in order to be indifferent between insuring or not. The optimal pool size is given by $m$, solution of the following equation:
\begin{equation}
\label{eq:insurance_pool}
    v(mP^+) = \mathbb{E}\left[v\left(mP^+ +P^+-\sum_{i=1}^N\min(\widetilde{Y_i},kw)\right)\right]. 
\end{equation}
As a comparison tool between insurer and insurance, we also compute the insurer relative wealth $w_i = mP^+/w$, which indicates how much bigger in terms of size, the insurer should be in order to issue the insurance contract to the company. Risk types with high relative wealth could be considered hard to insure, since the insurer would be required to have a much higher capital than the case of risk types with lower relative wealth. 

\begin{figure}[h]
	\centering
	{\includegraphics[width=\textwidth,height=0.8\columnwidth]{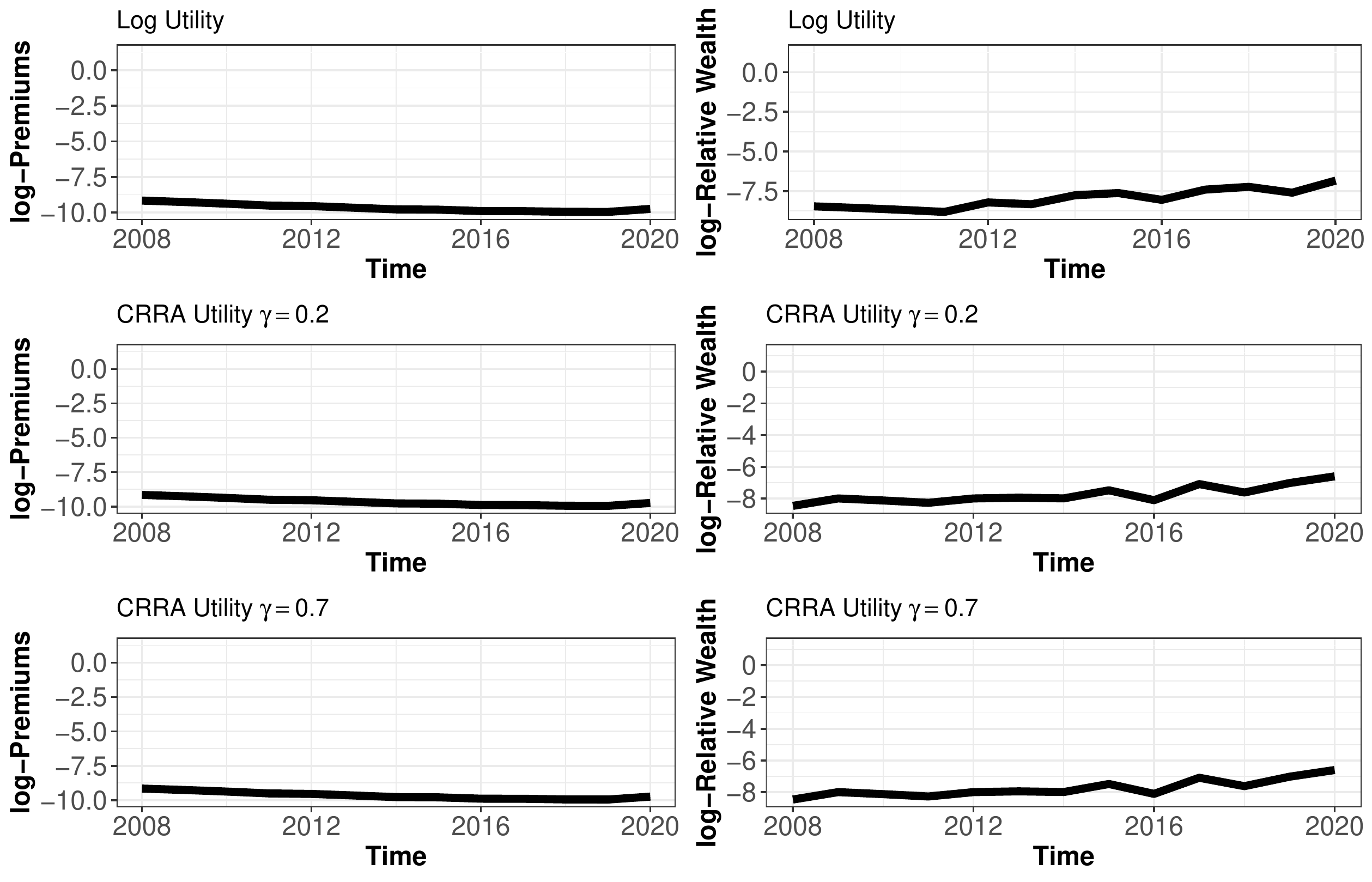}} 
	\caption[Insurable Case]{This figure shows premiums and the corresponding insurer relative wealth in million dollars, on the log scale, estimated using company specific characteristic, using logarithmic utility function, CRRA utility with $\gamma = 0.2$, and $\gamma= 0.7$, of ``Privacy - Unauthorized Contact or Disclosure". Company capital $w$ is equal to 1 billion dollars and the top cover limit $k$ is set to $10\%$ of company capital. The case depicted corresponds to a situation where the estimated tail parameter $\tau$ is greater than 1 and therefore the risk type appears to be insurable.}
	\label{Fig:premiums_insurable}
\end{figure}

\begin{figure}[h]
	\centering
	{\includegraphics[width=\textwidth,height=0.8\columnwidth]{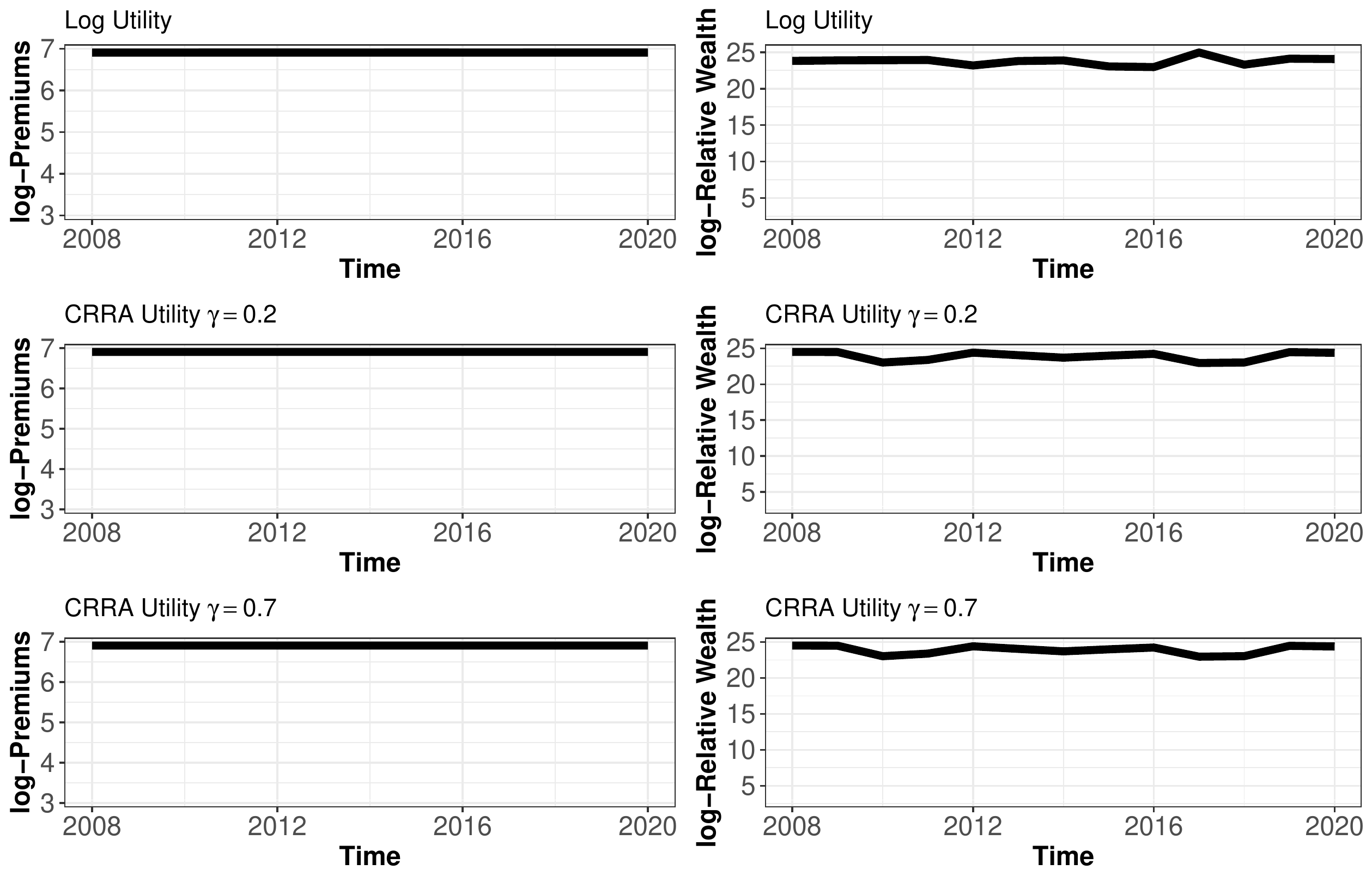}} 
	\caption[Too Expensive Case]{This figure shows premiums and the corresponding insurer relative wealth in million dollars, on the log scale, estimated using company specific characteristic, using logarithmic utility function, CRRA utility with $\gamma = 0.2$, and $\gamma= 0.7$, of ``Data Malicious Breach". Company capital $w$ is equal to 1 billion dollars and the top cover limit $k$ is set to $10\%$ of company capital. The case depicted corresponds to a situation where the estimated tail parameter $\tau$ is lower than 1 and therefore the risk type appears to be not insurable.}
	\label{Fig:premiums_tooexpensive}
\end{figure}

\begin{figure}[h]
	\centering
	{\includegraphics[width=\textwidth,height=0.8\columnwidth]{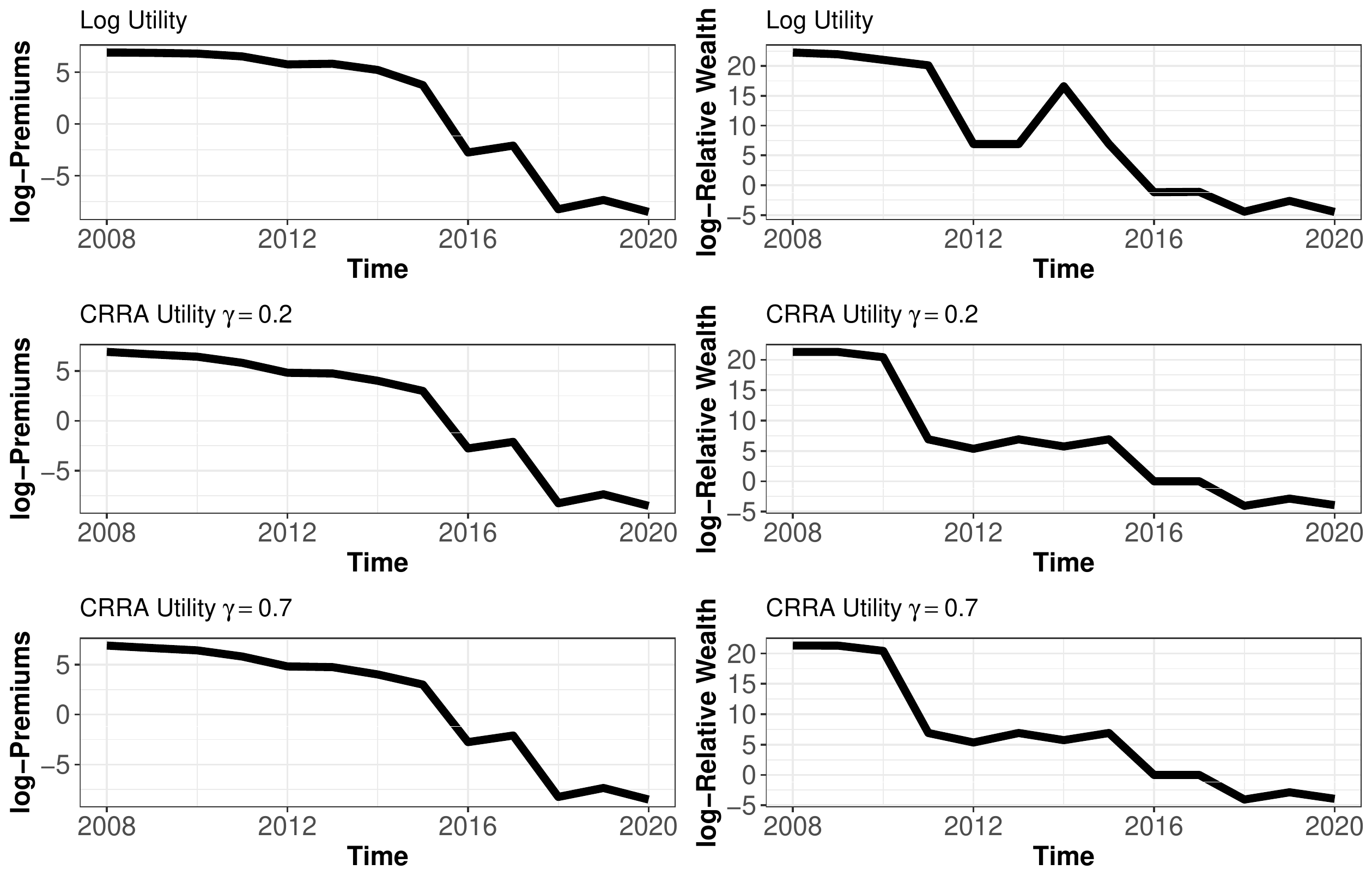}} 
	\caption[Not Manageable]{This figure shows premiums and the corresponding insurer relative wealth in million dollars, on the log scale, estimated using company specific characteristic, using logarithmic utility function, CRRA utility with $\gamma = 0.2$, and $\gamma= 0.7$, of ``Cyber Extortion". Company capital $w$ is equal to 1 billion dollars and the top cover limit $k$ is set to $10\%$ of company capital. The case depicted corresponds to a situation where the estimated tail parameter $\tau$ increases over time becoming greater than 1. The rapid change in the required wealth makes the risk type not insurable.}
	\label{Fig:premiums_notmana}
\end{figure}

The results of the combined POT and GAMLSS approach can be used to compute insurance premiums specific for a given company. This should help to address the insurabiltiy question at a company specific level, allowing to consider the heterogeneous nature of cyber risk and at the same time reducing uncertainty. We consider a financial company, residing in the USA, with high number of employees and high revenue, including also contagion type of events, and compute the insurance premiums for each risk types using a simulation approach.
Figures~\ref{Fig:premiums_insurable},~\ref{Fig:premiums_tooexpensive}, and~\ref{Fig:premiums_notmana} show the premiums and the corresponding insurer relative wealth in million dollars and on the log scale for ``Privacy - Unauthorized Contact or Disclosure", ``Data Malicious Breach", and ``Cyber Extortion" respectively, from 2008 to 2020, using logarithmic utility function, CRRA utility function with $\gamma = 0.2$, and $\gamma= 0.7$. Company capital $w$ is equal to 1 billion dollars and the top cover limit $k$ is set to $10\%$ of company capital. 

``Privacy - Unauthorized Contact or Disclosure" premiums and relative wealth in Figure~\ref{Fig:premiums_insurable} show the case when cyber risk is insurable, since both premiums and relative wealth are low. Nonetheless, this situations of low premiums is only attainable in the case of finite mean loss distribution, as shown in Figure~\ref{Fig:premiums_tooexpensive}. Figure~\ref{Fig:premiums_tooexpensive} shows the premiums and relative wealth of ``Data Malicious Breach", to which corresponds a case of infinite mean loss distribution. For the considered company, the computed insurance premium are too costly, making it  very hard, if not impossible to buy the insurance. The same applies to the offer side. In this case, the insurer requires to have a wealth more than $e^{20}$ times greater than the company total wealth. In cases where the underling cyber event severity distribution is of the infinite mean type, insurability is compromised.
Finally, Figure~\ref{Fig:premiums_notmana} depicts the case of ``Cyber Extortion" as an example of the effect of non stationary in the tail index on insurance premiums and relative wealth. As shown in Tables~\ref{table:model0_lambda},~\ref{table:model0_mu}, and~\ref{table:model0_sigma}, cyber event frequency and severity depend also on time. This time dependence also affect the insurability of cyber risk types, since the necessary wealth for the insurance company would need to be adjusted accordingly. The ``Cyber Extortion" depicted in Figure~\ref{Fig:premiums_notmana} shows that, while the premiums and relative wealth drop from non feasible to more feasible values, it also requires a rapid adjustment in the insurance company wealth, decreasing from being $e^{7}$ times higher the company wealth, to being a fraction of the company wealth in only 3 years.  
\clearpage

\section{Conclusions}\label{Sec:conclusions}
The expanding reliance of businesses and enterprises on information technology has resulted into an increase of the importance of cyber risk. Decision and policy makers have started to investigate the matter recently, and so does the actuarial community, with academic research, insurance industry, and risk managers. Nevertheless, the scarcity of good quality datasets is a common limitation among the many areas of study on cyber risk. We used the industry leading dataset provided by Advisen, to study cyber risk modelling and insurance aspects. In particular, we focused on two main unresolved questions on cyber risk: which factors are important explanatory variables for cyber event frequency and severity, and what can be inferred on cyber risk insurability. 

In our analysis, we found that cyber event severity distribution is of the infinite mean type, having important implications both from modelling and insurance perspective. Regarding the modelling of cyber risk, we found that no standard OLS based techniques would be able to estimate correctly relevant parameters. We then used the POT method combined with the GAMLSS approach allowing cyber event frequency and severity distributional parameters to depend on covariates. Using the Voung's closeness test \citep{vuong1989} we found that joint GAMLSS regression structure across all cyber risk types is not statistically distinguishable from a separate estimation of a GAMLSS for each risk type. Which implies that distinguish statistical attributes of cyber risk types, important from a insurance perspective such as the tail behavior, it is not possible. Then, combining the POT method with the GAMLSS regression approach we captured the complex and heterogeneous nature of cyber risk, showing  that time trends, and the impact of company size, business sector, and contagion on cyber event frequency and severity varies with cyber risk type. A further investigation using a ordinal rank regression framework confirmed that the results are likely to be mainly affected by extreme event, regardless of the cyber risk type under consideration.

Our extensive empirical analysis allowed us to translate statistical features of cyber event frequency and severity to capital requirements and cyber risk insurability. We showed how the inclusion of covariates in cyber risk modelling can improve capital requirement calculations, allowing for company and risk management tailored approaches for cyber risk types. Finally, we discussed the implications of our statistical findings on insurance premium calculation and more broadly, cyber risk insurabiltiy. Using a utility based argument, we show that consistently with the findings of the empirical analysis, insurance premiums vary with cyber risk types.  According to our  calculations, based on the combined POT method and GAMLSS approach, tail behavior and non stationarity in the tail parameter could jeopardize both demand and offer for cyber risk insurance, to the point where insurance premiums and insurance pool would be too high to be realistically adopted, or require a excessively high level of active management that seldom insurance company exhibit or have incentive to undertake. 

Our study is one of first to thoroughly investigate statistical features of monetary losses related to cyber risk. Our findings provide useful insights for industry and market participants, as well as policy makers and regulators.

\section{Acknowledgement}
This research has been conducted within the Optus Macquarie University Cyber Security Hub and funded by its Risk Management, Governance and Control Program.
We would like to acknowledge valuable comments from participants at the international congress ``Insurance: Mathematics and Economics" 2021.

~

%

\section{Appendix}
\subsection{Goodness of Fit}
\label{appendix:ks}
Table~\ref{table:ks.1} shows log-likelihood, Akaike information criterion (AIC) and p-values for the Kolmogorov\--Smirnov (KS) test for some commonly used distributions for cyber event severity. The null hypothesis is rejected for every considered distribution (the null hypothesis being no difference between cyber related loss distribution and one of the considered distribution), suggesting that more complex estimation procedures, such as the combined POT and GAMLSS approaches, may be required. Among the chosen distributions, lognormal, generalized Pareto, and log-logistic preform better in terms of AIC and log-likelihood. To allow for heterogeneity in cyber event severity, we also perform the KS test on cyber event severity by cyber risk types.

\begin{table}[ht!]
	\centering
	\caption{This table reports the log-likelihood values, AIC and KS p-values for different distributional choices for cyber event severity. The null hypothesis is rejected for all the considered distributions.}
	\label{table:ks.1}
	\begin{tabular}{l|rrr}
		Distribution &	Log-Likelihood	& AIC	& KS test	\\
		\hline
		
     Exponential &	-30,211.2855& 	60,426	&0\\
     Gamma &	0	 &6	&0\\
     Generalized Pareto	&-8,091.2171 &	16,188 &	0\\
     Log-logistic	&-8,186.3720&	16,378&	0\\
     Lognormal&	-7,825.6678&	15,657&	0\\
     Weibull&	-10,350.5367&	20,707	&0\\
     skew-Normal	&-40,442.5800&	80,891	&0\\

		\hline	
		
	\end{tabular}
\end{table}

Table~\ref{table:ks.2} shows the p-values for the KS test for some commonly used distributions for cyber event severity, broken down by cyber risk types. With the exception of cyber related events of the type ``Privacy - Unauthorized Contact or Disclosure'', for losses belonging to any other cyber risk type it's not possible to reject the null hypothesis of being distributed as, generalized Pareto, lognormal or log-logistic. 
\begin{table}[ht!]
	\centering
	\caption{This table reports the p-values for a KS test for different distributions broken down by risk types. Except for ``Privacy - Unauthorized Contact or Disclosure", every other cyber risk type can be fitted on a generalized Pareto, lognormal, or log-logistic distribution.}
	\label{table:ks.2}
	\begin{adjustbox}{width=\columnwidth}
	\begin{tabular}{lccccccc}
	Risk Type &	Exponential &	Gamma &	GPD & Log-logistic& Lognormal & Weibull &	skew-Normal\\
	\hline 
	
	Privacy - Unauthorized Contact or Disclosure 	&0   &	0	&    0   &	0&	0&	0&	0\\
	Privacy – Unauthorized Data Collection 	&0   &	0   &	0.41 &	0.69 &0.16 &	0&	0\\
	Data - Physically Lost or Stolen&0   &  0   &	0.31 &	0.16 &	0.24&	0&	0\\
	Identity – Fraudulent Use/Account Access	    &0   &  0   &	0&	0.54&	0.85&	0&	0\\	
	Data - Malicious Breach			&0   &	0   &	0&	0	&0.92&	0&	0\\
	Phishing, Spoofing, Social Engineering 			&0   &	0   &	0.60&	0.16	&0.06&	0&	0\\
	IT -  Configuration/Implementation Errors				&0   &	0.02&   0.28 &	0.24&	0&	0&	0\\
	Data - Unintentional Disclosure &0   &	0   &	0.14&	0.55&	0.43&	0&	0\\
	Cyber Extortion					&0   &	0   &	0.06 & 0.001&	0.83&	0&	0\\
	Network/Website Disruption		&0   &	0   &	0.79 &	0.12&	0.49&	0&	0\\
	Skimming, Physical Tampering	&0   &	0	&   0.60 &	0.27&	0.58&	0&	0\\
	IT - Processing Errors			&0   &	0	&   0.15 &	0.08	&0&	0.01&	0\\
	Industrial Controls 			&0.02&	0.89&	0    & 0.70&	0.22	&0.44&	0\\
	Undetermined/Other				&0.22&	0.75&   0.94&	0.45&	0.45&	0.20&	0.01\\
		
	\hline 
		
	\end{tabular}
	\end{adjustbox}
\end{table}
These results are in line with the findings in the literature \cite[see, e.g.][]{edwards2016, eling2017data, eling2019actual}, where lognormal, generalized Pareto, and log-logistic distributions are often considered valid alternatives for cyber event severity. The implication of these results on the goodness of fit of the GAMLSS approach is investigated in Section~\ref{appendix:gpd_vs_lognormal}.
\subsection{Non-Linearity and Interactions}
\label{appendix:non_linearity_interaction}
Dynamic EVT allows for a flexible structure in the link functions, including parametric, semi-parametric and non parametric relationship. We follow the spline smoothing approach in \cite{chavez2016} and select the values of $\gamma_\lambda,\gamma_\mu$ and $\gamma_\tau$ according to the AIC \cite[see, also][]{ganegoda2013,eling2019actual}. 
\begin{figure}
    \centering
    \includegraphics[width=0.8\textwidth,height=0.4\columnwidth]{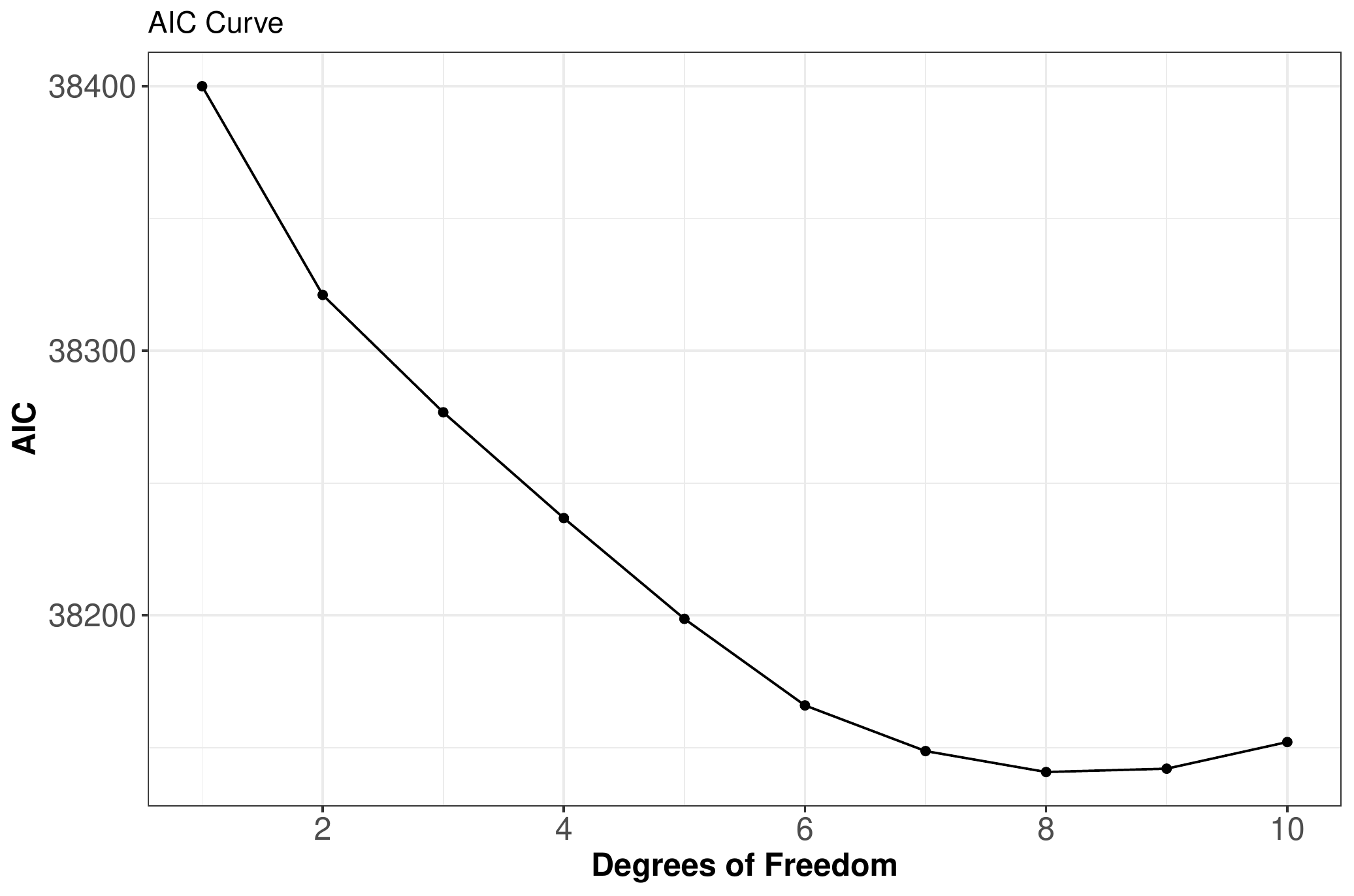}
    \caption{AIC curve for cyber event frequency for different values of $\gamma_\lambda$. $\gamma_\lambda = 8$ corresponds to the minimum value of AIC.}
    \label{fig:dof_frequency}
\end{figure}
Figure~\ref{fig:dof_frequency} shows the AIC curve for different values of degrees of freedom in the smoothing spline in Equation (\ref{eq:link}). The minimum AIC is reached at $\gamma_\lambda$ equals to $8$, suggesting that cyber event frequency depends non linearly on time.
\begin{figure}
    \centering
    \includegraphics[scale=0.6]{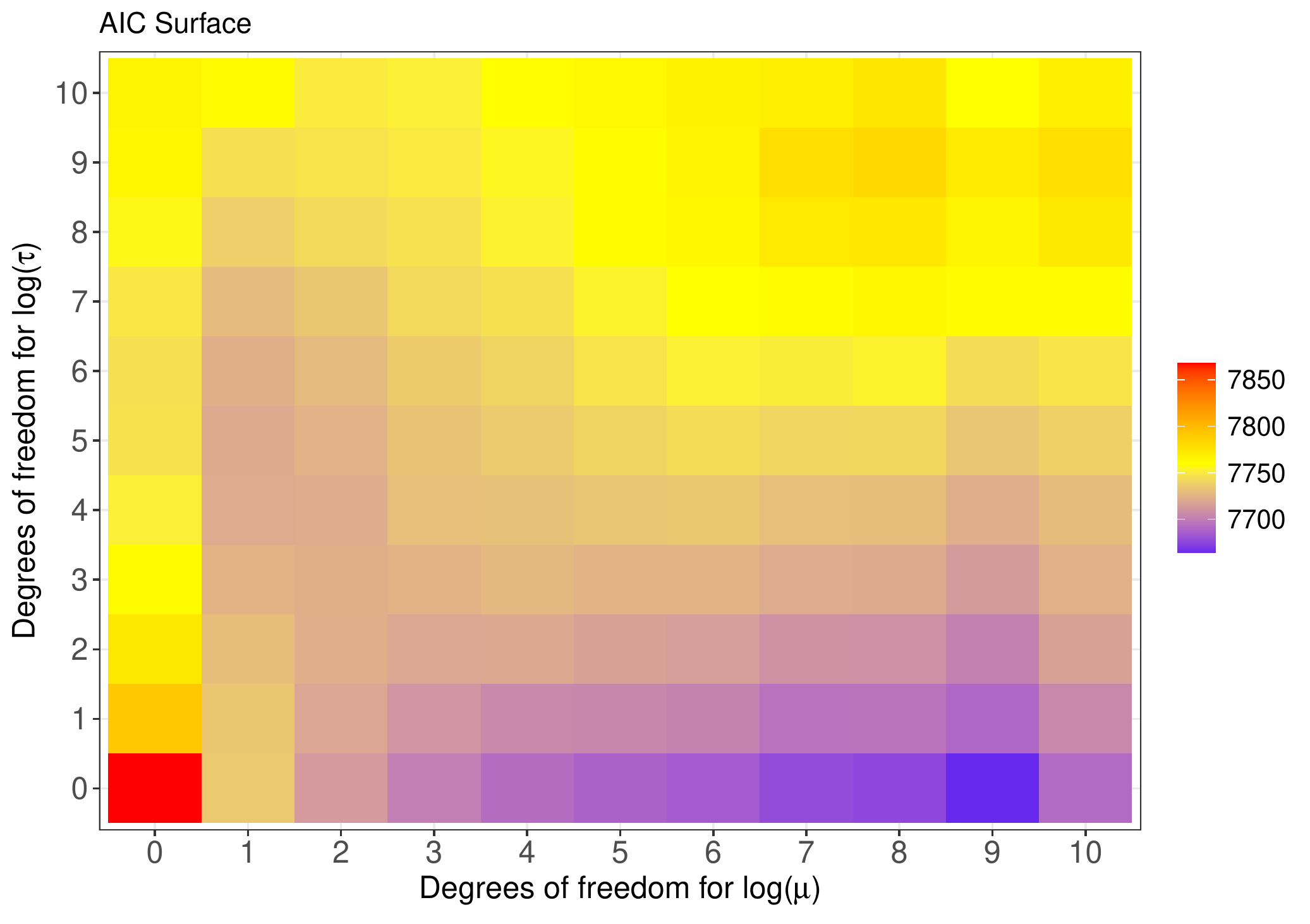}
    \caption{AIC surface for cyber event severity for varying values of $\gamma_\mu$ and $\gamma_\tau$. The minimum value of the AIC suggests the tail parameter $\tau$ should not depend on time and a non-linear relationship between $\log(\mu)$ and time.}
    \label{fig:dof_severity}
\end{figure}

Figure~\ref{fig:dof_severity} shows the AIC surface for different values of degrees of freedom in the smoothing splines in the case of $\mu$ and $\tau$. The minimum values of the AIC suggests that the tail parameter $\tau$ should not depend on time, while $\log(\mu)$ appears to depend non linearly on time.

\begin{figure}
    \centering
    \includegraphics[scale=0.6]{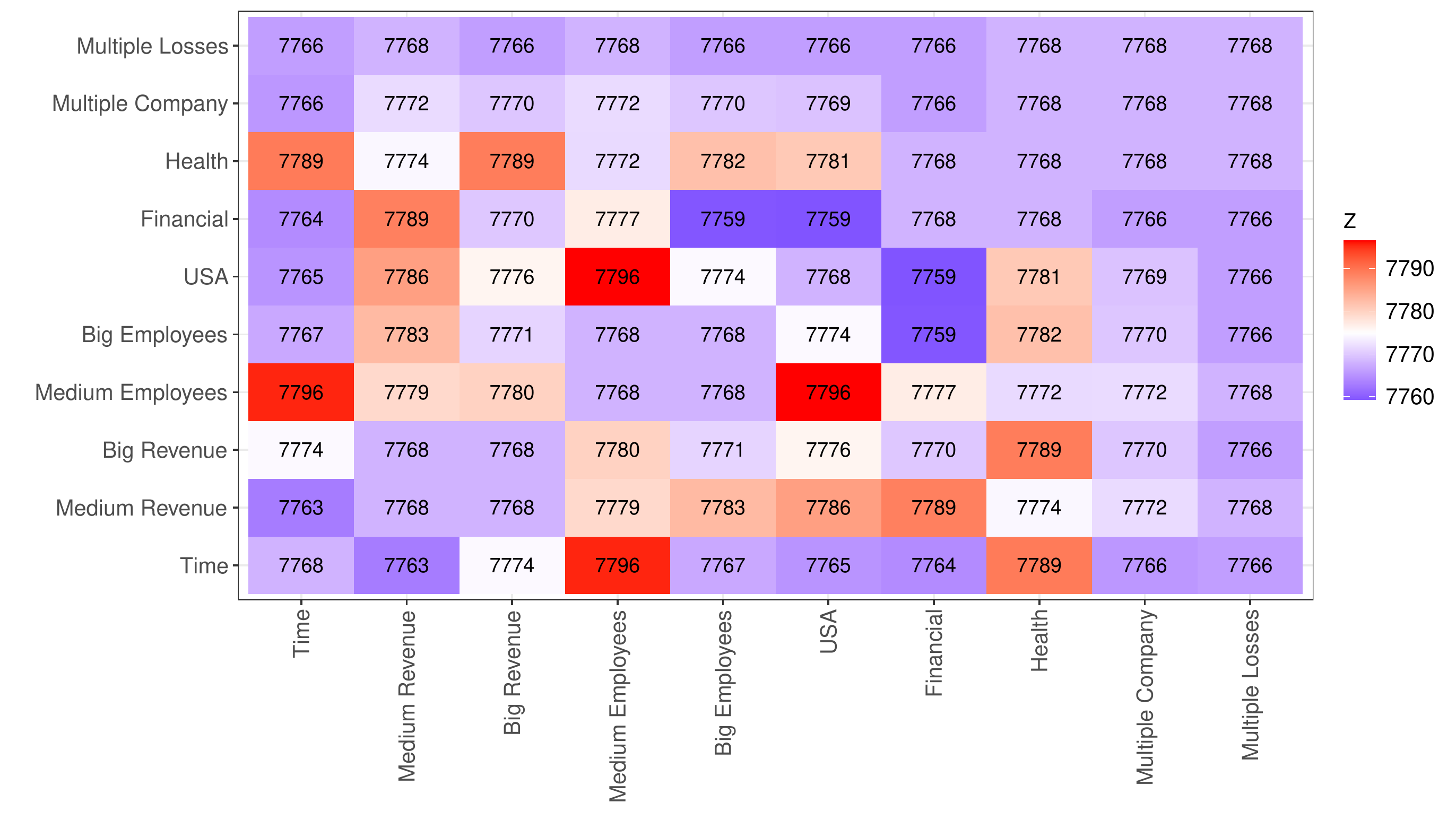}
    \caption{AIC values for models including different interaction terms between covariates.}
    \label{fig:aic_interaction}
\end{figure}

\begin{figure}[ht]
	\centering
	\subfloat[{Q-Q plot of residual of model without interaction term }]{{
	\includegraphics[width=0.45\textwidth,height=0.3\columnwidth]{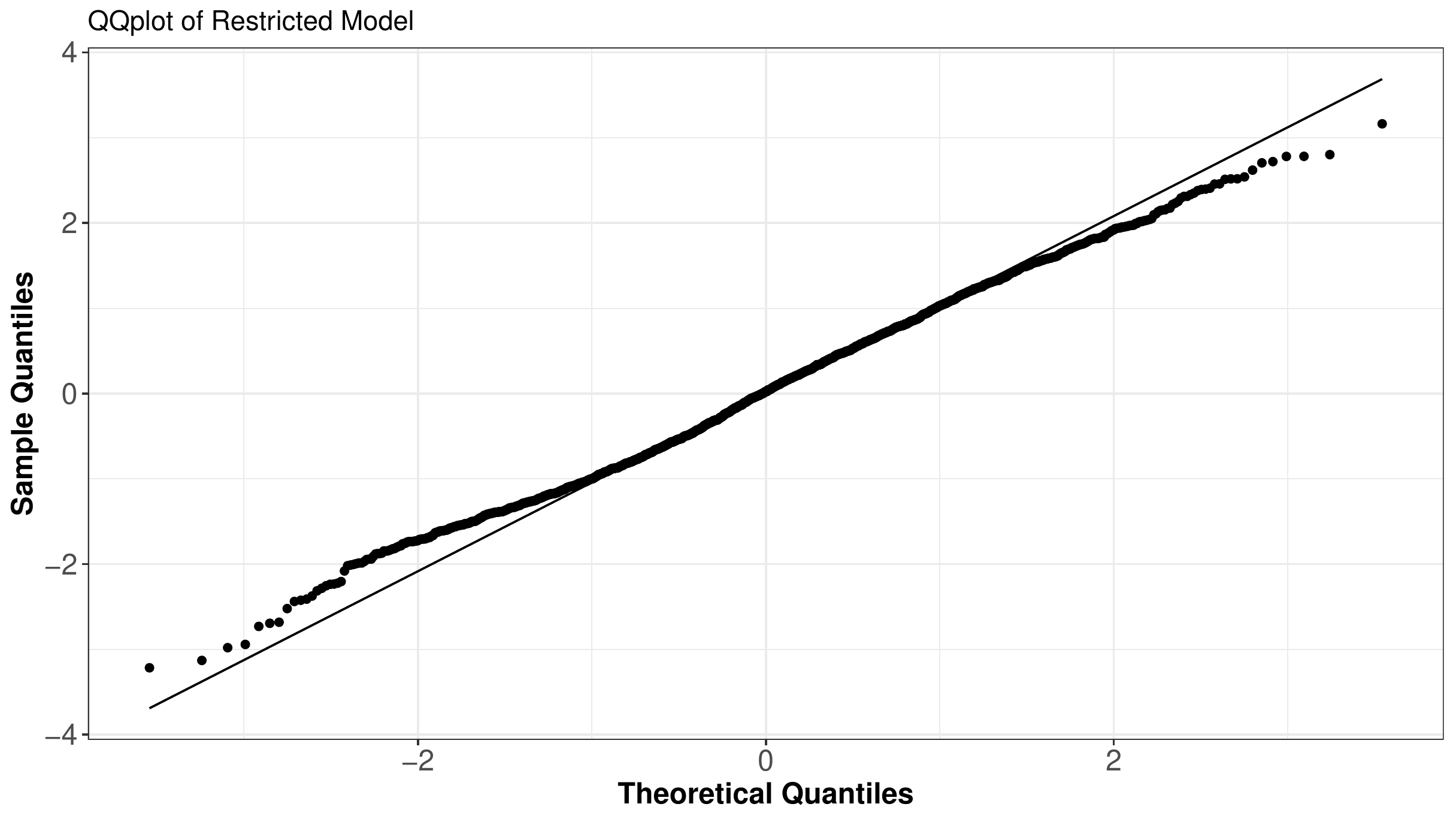}}}
	\qquad
	\subfloat[{Q-Q plot of residual of model with interaction term between time and company size}]{{\includegraphics[width=0.45\textwidth,height=0.3\columnwidth]{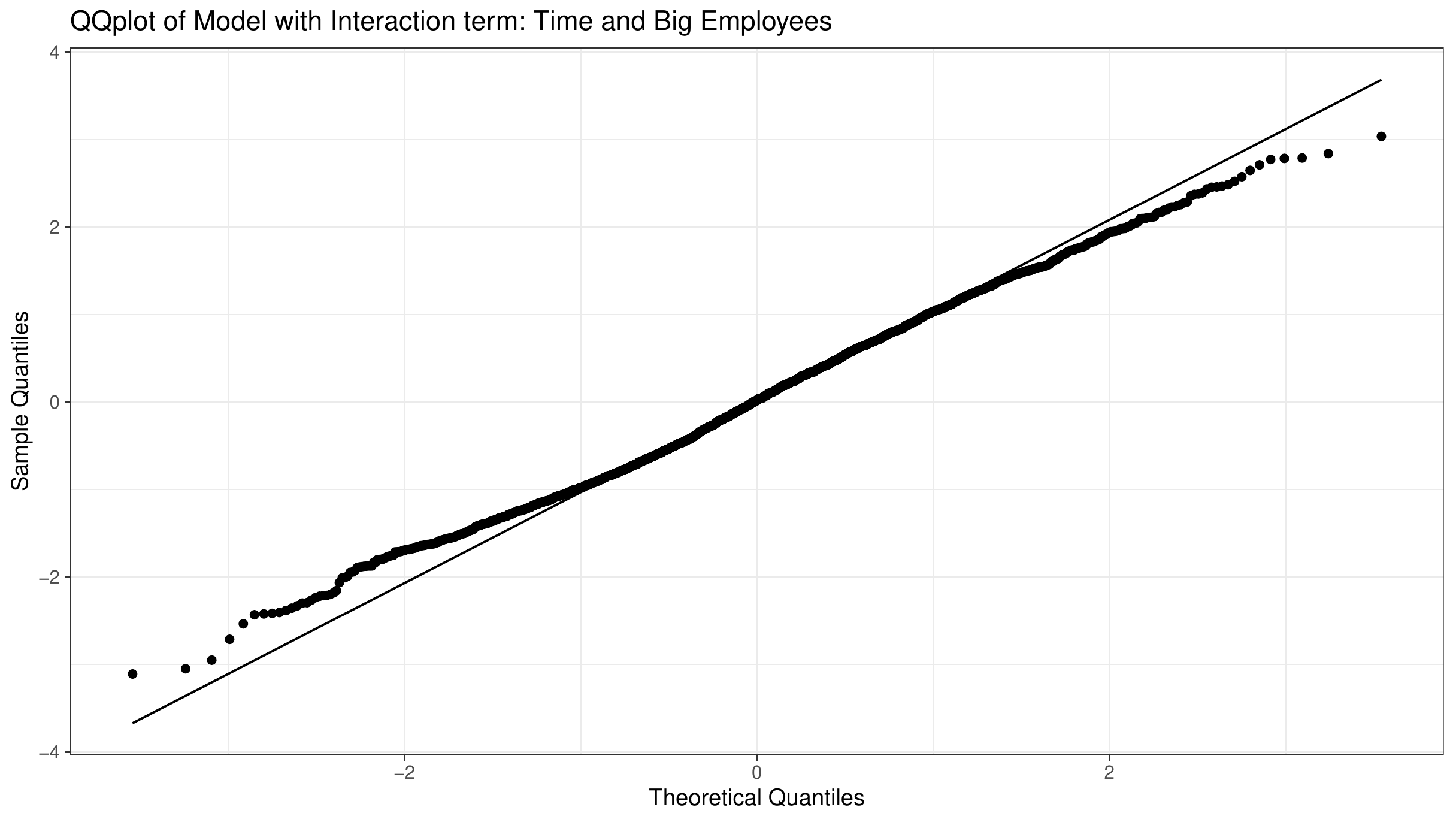}}}
	\qquad
	\subfloat[{Q-Q plot of residual of model with interaction term between USA and Finance}]{{\includegraphics[width=0.45\textwidth,height=0.3\columnwidth]{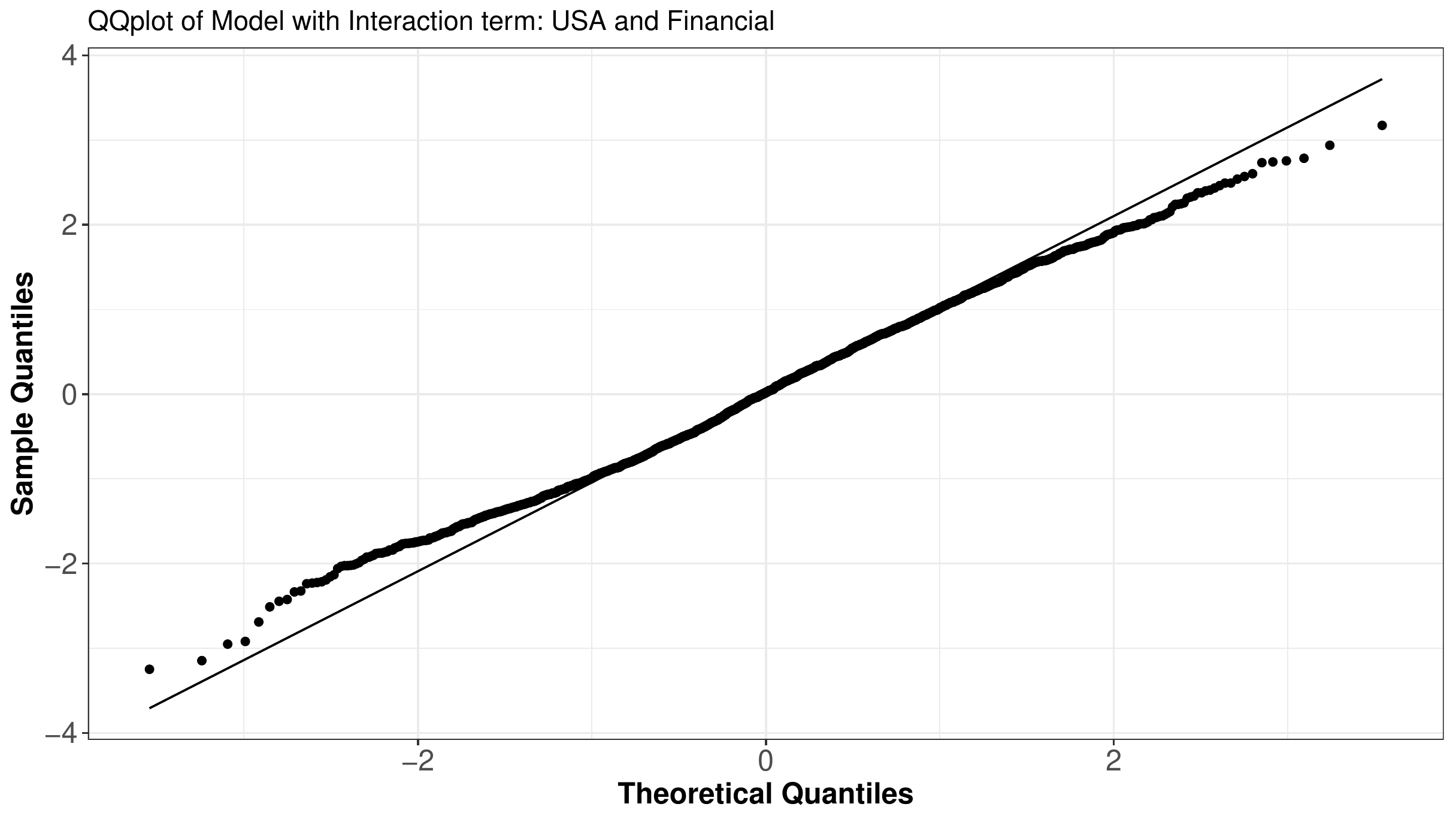}}}
	\caption[Q-Q plot Interaction]{This figure shows the Q-Q plots in three different configurations: no interaction term in the link functions (panel (a)), interaction between time and size (panel (b)), and interaction between Finance and USA (panel (c)).}
	\label{Fig:qqplot_interaction}
\end{figure}

We also use the AIC to investigate if any interaction effect among covariates is present, with particular interest in the effects on severity. For any possible combination, we include the interaction term as covariate in both $\log(\mu)$ and $\log(\tau)$ and estimate the corresponding model. Figure~\ref{fig:aic_interaction} shows changes in the AIC surface when interaction terms are included. The values on the main diagonal correspond to the no interaction term case. 
As it can be seen from Figure~\ref{fig:aic_interaction}, some interaction terms show improvements in the AIC values with respect to the restricted model. Nonetheless, the impact of including any interaction term on the goodness of fit is minimal. To illustrate this, we compare the residual Q-Q plot of the model without interaction term and two interaction cases: interaction term between USA and Financial (with an AIC equal to 7759), and interaction term between time and company size (with an AIC equal to 7767). Figure~\ref{Fig:qqplot_interaction} shows the residual Q-Q plot of cyber event severity fitted in three different configurations: no interaction term in the link functions (panel (a)), interaction between time and size (panel (b)), and interaction between Finance and USA dummy covariates (panel (c)). No particular difference arises from a graphical inspection of the three selected alternatives, implying that while adding an interaction term might improve a information criterion, it does not necessarily increases the goodness of fit.

\clearpage
\subsection{$RGA_R$ based testing methodology}
\label{appendix:rga_test}
In the context of ordinal regression, a formal hypothesis test to evaluate two competing models can be constructed using the following test statistic \cite[see,][]{giudici2020cyber}:
\begin{equation}
    T = n\Bar{r}\left(RGA_{R_{full}} - RGA_{R_{rest}}\right),\notag
\end{equation}
where $\Bar{r}$ is the average rank, $RGA_{R_{full}}$ and $RGA_{R_{rest}}$ are the $RGA_R$ of the full and the restricted model, respectively. The test statistic $T$ is then distributed as variance gamma distribution, with parameters $\lambda = n/2$, $\alpha = 1/2$, $\beta = 0$, and $\mu = 0$, with $\lambda>0$, $\alpha\in\mathbb{R}$, $\beta$ is the asymmetry parameter, and $\mu$ is the location parameter. Given the fact that, for large values of $\lambda$ the distribution of the test statistic does not well behave, \cite{raffinetti2015} suggests to utilize $d$ sub-samples to robustify the test statistics and compute the significance values as follows:
\begin{equation*}
    s\text{-value} = \mathbb{P}\left[T \geq |t_{\alpha/2}| \right] = \frac{1}{d}\sum_{i=1}^d\mathbb{I}_{T \geq |t_{\alpha/2}|},
\end{equation*}
where $\mathbb{I}_{T \geq |t_{\alpha/2}|}$ is equal to 1 whenever $T \geq |t_{\alpha/2}|$ and 0 otherwise\footnote{It is important to notice that the $s$-value is not to be interpreted as a p-value, but rather as a test statistics for the proportion of times the value of T is greater than a given threshold in the subsamples. The authors themselves suggest to perform a further hypothesis test on the $s$-value based on the binomial distribution \cite[see,][]{raffinetti2015}.}. Following \cite{raffinetti2015}, the quantity $\sum_{i=1}^d\mathbb{I}_{T \geq |t_{\alpha/2}|}$ follows a binomial distribution with parameters $d$ and $p$, with the following probability mass function:
\begin{equation}
    f\left(n; d, p \right) = \binom{d}{n} p^n(1-p)^n 
\end{equation}
with $n=1,\dots,d$. In other words, $s$-value is an estimator for the quantity $p$ to which corresponds the probability of success in the binomial distribution. To test whether the estimates for $s$-value are consistent with every possible subsampling combination, a three way testing procedure can be used. Call $z =  (\widehat{s\text{-value}}-p_0)/\sqrt{p_0(1-p_0)/d} $
\begin{itemize}
    \item If $\widehat{s\text{-value}}\in(0.7,1]$, perform $H_0: p\leq 0.7$ vs $H_1: p>0.7$. Reject the null if $\mathbb{P}[Z>z]\leq \alpha_s$
    \item If $\widehat{s\text{-value}}\in(0.0,0.3]$, perform $H_0: p> 0.3$ vs $H_1: p\leq0.3$. Reject the null if $\mathbb{P}[Z\leq z]\leq \alpha_s$
    \item $\widehat{s\text{-value}}\in(0.3,0.5]$, two tests need to be performed: $H_0: p\leq 0.3 $ vs $H_1: p> 0.3$ and, $H_0: p>0.5$ vs $H_1:p\leq 0.5$. Then the p-values of both test are combined using Holm's two steps approach \cite{holm1979}. The case of $\widehat{s\text{-value}}\in(0.5,0.7]$ follows analogously.
\end{itemize}
    
Table~\ref{table:rank_regression} shows the result on the significance of each individual covariates in the linear model in Equation (\ref{eq:rank_regression}), using the Advisen classification.      
 \begin{table}[h]
	\centering
	\caption{This tables show the results of the rank based regression analysis on model 0. The $s$-values corresponds to the test of hypothesis for individual significance of each coefficient. The size of the sub-sample is set equal to 10, and the number of sub-samples drawn is 5,000. None of the Advisen risk types are statistically significant}
	\label{table:rank_regression}
	\begin{adjustbox}{width=\columnwidth}
	\begin{tabular}{l|lllll}
		Covariates                      & RGA     & $s$-value & $p_s$-value & Result   & s-calss\\
		\hline
        Privacy - Unauthorized Contact or Disclosure  & 0.2168 & 0       &     -      &     -     & Never significant \\
        Data - Unintentional Disclosure & 0.2168 & 0       &     -      &     -     & Never significant\\
        Privacy – Unauthorized Data Collection     & 0.2168 & 0       &     -      &     -     & Never significant\\
        Data - Malicious Breach         & 0.2168 & 0       &     -      &     -     & Never significant\\
        Identity – Fraudulent Use/Account Access        & 0.2168 & 0       &     -      &     -     & Never significant\\
        Data - Physically Lost or Stolen  & 0.2168 & 0       &     -      &     -     & Never significant\\
        Skimming, Physical Tampering & 0.2168 & 0       &     -      &     -     & Never significant\\
        IT - Processing Errors          & 0.2168 & 0       &     -      &     -     & Never significant\\
        Phishing, Spoofing, Social Engineering         & 0.2168 & 0       &     -      &     -     & Never significant\\
        IT - Configuration/Implementation Errors  & 0.2168 & 0 &     -      &     -     & Never significant\\
        Network/Website Disruption      & 0.2168 & 0       &     -      &     -     & Never significant\\
        Cyber Extortion                 & 0.2168 & 0       &     -      &     -     & Never significant\\
        Undetermined/Other              & 0.2168 & 0       &     -      &     -     & Never significant\\
        Industrial Controls/Operations  & 0.2168 & 0       &     -      &     -     & Never significant\\
        Time                            & 0.1861 & 0.9002  &  $<0.0001$ &     $p > 0.7$     & Almost Always Significant\\
        $R_{medium}$                      & 0.2094 & 0.5344  &    0       &     $p > 0.3$     & Frequently significant\\
        $R_{big}$                         & 0.2190 & 0.2112  &    0.0633  &     $p > 0.3$    & Rarely significant\\
        $E_{medium}$                      & 0.2051 & 0.5852  &    0.0022  &     $p \in[0.5,0.7]$    & Frequently significant\\
        $E_{big}$                         & 0.1881 & 0.8130   &    0       &     $p > 0.7$     & Almost Always Significant\\
        $L_{USA}$                             & 0.2176 & 0.7322  &    0.2563  &     $p \leq 0.7$     & Almost Always Significant\\
        $B_{financial}$                       & 0.2154 & 0.3042  &    0.4136  &     $p > 0.3$     & Sometimes significant\\
        $B_{health}$                          & 0.2164 & 0.0722  &  $<0.0001$ &     $p \leq 0.3$     & Rarely significant\\
        MC                & 0.2165 & 0.0386  &  $<0.0001$ &     $p \leq 0.3$     & Rarely significant\\
        ML                 & 0.2167 & 0.0172  &  $<0.0001$ &     $p \leq 0.3$     & Rarely significant\\

		\hline	
	\end{tabular}
	\end{adjustbox}
\end{table}
For each covariate Table~\ref{table:rank_regression} shows the  RGA, $s$-value, and the $s$-class of the restricted model. The $s$-class is computed accordingly to \cite{raffinetti2015} and \cite{giudici2020cyber}. As it can be seen from the table, only Time, $E_{big}$ and $L_{USA}$ belong to the significance class ``Almost Always" significant, while none of the Advisen Risk types are statistically significant. To further investigate this we also have run the test for the joint significance of the risk types getting a $s$-value of 0.4982.

\subsection{Coupled Model Estimates}
\label{appendix:gpd_vs_lognormal}
In this section we present the dynamic EVT estimates for the coupled model for cyber severity. We proceed with the approach discussed in Section~\ref{sec:model}, finding that the threshold values for the coupled model is $59\%$, similarly the results discovered in \cite{eling2019actual}. Then, for cyber related losses exceeding such threshold we fit the model given by likelihood and link functions in Equation~(\ref{eq:joint_model}). Given the results in Section \ref{appendix:ks} of the KS test, we consider also two alternative distributions, commonly used in Operational Risk applications, log-logistic and lognormal. Table~\ref{table:5Jointly} shows the estimates for each distribution. Comparing the results for the Generalized Pareto distribution with the results of the decoupled model it can be seen that fewer  coefficients appear to be statistically significant. Looking at the tail parameter $\tau$, it shows that a company with a high level of revenue tends to experience heavier cyber related losses, while for a company residing in the USA the severity seems to be lower than that for companies residing outside the USA. Indicating that perhaps cyber awareness, mitigation or risk management processes are more complete in companies located in the USA versus equivalent entities outside the USA. Similar consideration can be made for companies operating in the healthcare or financial business sectors: cyber event losses appears to be less severe in financial or healthcare companies with respect to enterprises operating in other business sectors. This can once again be explained by the higher degree of cyber awareness present in these industries. Looking at the temporal dependence, the evidence appears to be mixed, with lognormal and generalized Pareto distributions returning conflicting, although not statistically significant, results. Looking at the effect of cyber risk types, very few coefficients are statistically significant, in line with the previous results obtained in Section~\ref{Sec:ranked_based_regression}. Overall, when the coupled model is considered, cyber event severity appears to be mainly driven by company specific characteristics, while cyber risk types, time, and contagion are only rarely significant variables. 

\begin{figure}[ht]
    \centering
    \includegraphics[width=0.95\textwidth,height=0.6\columnwidth]{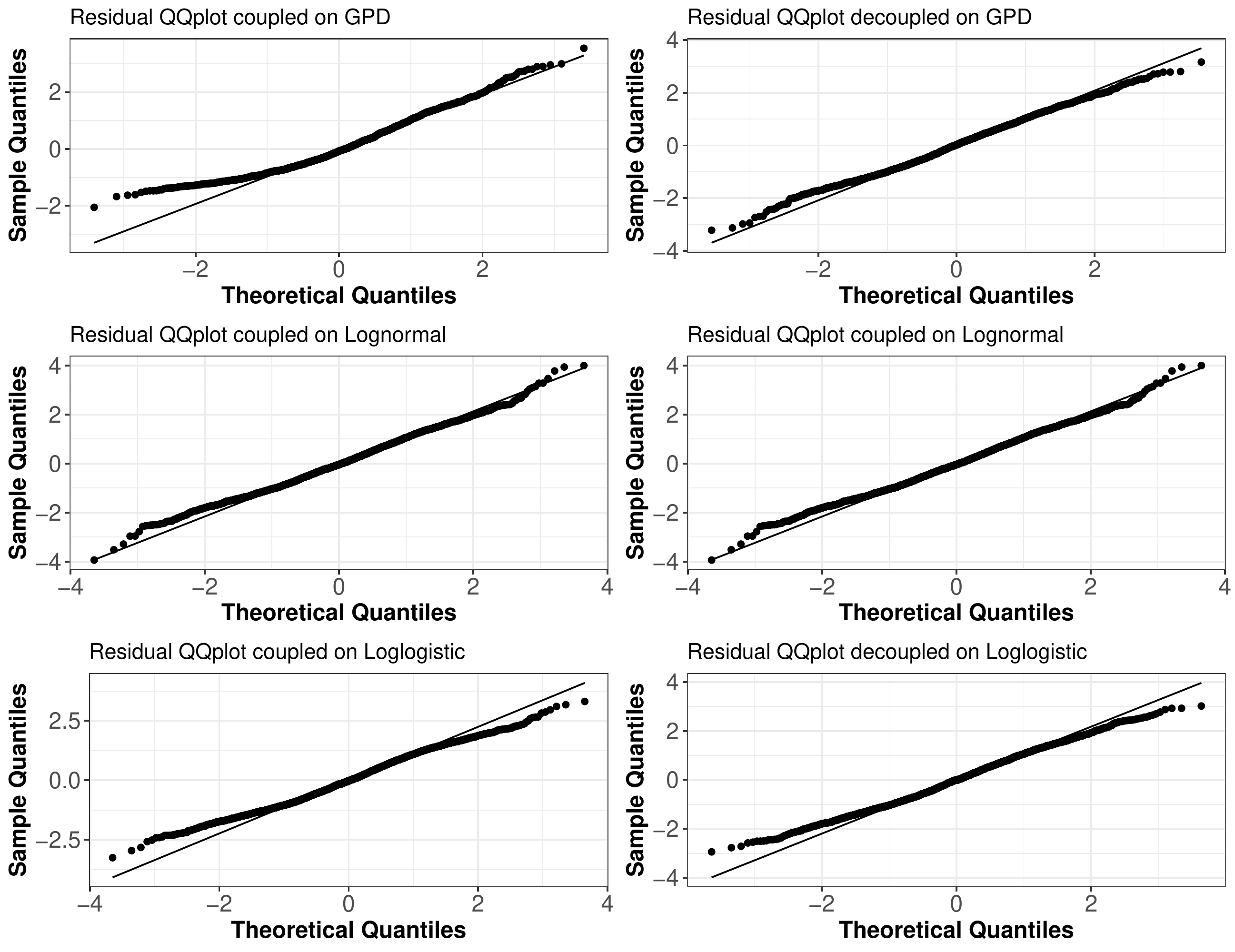}
    \caption{This figure shows the residual Q-Q plots of cyber event severity fitted on generalized Pareto, lognormal, and log-logistic  distributions coupled (on the left) and decoupled (on the right).}
    \label{Fig:qqplot_alternative}
\end{figure}

\begin{table}[ht!]
	\centering
	\caption{Coefficient Estimates for three different severity loss distribution regression fits under the GAMLSS setting. Note: the threshold for the GAMLSS GPD regression model was selected at 59\%}
	\label{table:5Jointly}
	\begin{adjustbox}{width=\columnwidth}
	\begin{tabular}{lllllll}
		Covariates & \multicolumn{2}{c}{logNormal}& \multicolumn{2}{c}{GPD}& \multicolumn{2}{c}{logLogistic}\\
		&    $\mu$ & $ \log(\sigma)$ & $\log(\mu)$ & $\log(\tau)$ & $\mu$ & $ \log(\sigma)$\\
	\hline 
		$\beta_0$                      &-316.6092***&	0.5983    & -63.8505*     &	-17.9724  &-340.8670*** &	0.5983***\\
		Time	                       &0.1647***   &	0.0001   &	0.0407**      &	0.0095   &  0.1769***&  0.0001***\\
		$R_{medium}$                   &-0.2223     &	-0.0750*  &	-0.3536**     &	-0.1263  & -0.2655*  &	-0.0750*\\
		$R_{big}$	                   &0.5367**    &	0.0055    &	0.0435         &	-0.2623**&  0.5353***&	0.0055\\
		$E_{medium}$                   &0.2635 *    &	0.0387    &	0.1971         & 0.1166   & 0.2894*   &	0.0387\\
		$E_{big}$	                   &1.0606***   &	0.0315    &	0.4278**       &	0.1440   & 1.1461*** &	0.0315\\
		$L_{USA}$	                   &0.7166***   &	-0.0465   &	0.6562***      &	0.3093***& 0.7710*** &	-0.0465\\
		$B_{financial}$                &0.2708*     &	0.0144*** &	0.6492***      &	0.3318***& 0.3368*   &	0.0144\\
		$B_{health}$	               &0.0235      &	-0.2480    &	0.4185**   &	0.7531***& 0.0625    &	-0.2480***\\
		ML                             &-2.0848     &	-0.1139    &	-0.5483**  &	0.9510   & -2.3437   &	-0.1139\\
		MC                             &-3.2110     &	-0.1313    &	-4.7323*** &	-1.2791* & -3.8585*  &	-0.1313\\
		Privacy - Unauthorized Contact or Disclosure   &-1.6194**   &	0.3844*    &	1.2107**   &	0.0272   & -1.6418** &	0.3844\\
		Privacy – Unauthorized Data Collection      &-0.1683     &	0.2438     &	-0.1922    & -0.5236  &  0.0014   &	0.2438\\
		Data - Physically Lost or Stolen  &-0.7374     &	0.3264     &  -0.6194      &	-0.7809* &  -0.6587  &	0.3264\\
		Identity – Fraudulent Use/Account Access  	   &-2.5693***  &	0.2877    &	-0.3174        &	-0.2325  & -2.4116***&	0.2877\\
		Data - Malicious Breach          &-0.1813     &	0.2428    &	-0.0340        &	-0.5986  & -0.0989   &	0.2428\\
		Phishing, Spoofing, Social Engineering	           &-0.3590     &	0.0393    &	-0.3976        &	-0.4555  & -0.2996   &	0.0393\\
		IT -  Configuration/Implementation Errors  	           &0.1785      &	0.2843    &	0.0494         &	-0.7946* & 0.2943    &	0.2843\\
		Data Unintentional Disclosure  &-1.2169*    &	0.0559    &	0.2962         &	0.1148   & -1.0761   &	0.0559\\
		Cyber Extortion 			   &-4.0761***  &   0.0429   &	-0.7456        &	-0.2864  & -3.9993*** &	0.0429\\
		Network/Website Disruption	   &-0.8125     &	0.3320    &	-0.2308        &	-0.8477**& -0.8070   &	0.3320\\
		Skimming, Physical Tampering	   &-2.1929     &	0.2582    &	-0.1169        &	-0.1578  & -2.1180** &	0.2582\\
		IT – Processing Errors	       &0.6990***   &	0.4725*   &	1.1321         &	-0.8106* & 0.8234   &	0.4725\\
		Industrial Controls &0.4209      &	0.2258    &	0.8873         &	-0.5504  & 0.3442   &	0.2258\\

	\hline 
		
	\end{tabular}
	\end{adjustbox}
\end{table}
Finally, Figure~\ref{Fig:qqplot_alternative} compares Q-Q plots of cyber event severity fitted on generalized Pareto, lognormal, and log-logistic  distributions coupled (on the left) and decoupled (on the right). Both in the case of coupled and decoupled models, the Q-Q plots look very similar for all the distributions. This fact has a twofold implication. First, it further confirms that, coupled and decoupled models for cyber event severity are statistically indistinguishable. Second, while the combined POT and GAMLSS approach is a powerful but sophisticated framework, other commonly used distribution such as log-logistic and lognormal performs adequately well in fitting cyber event severity. 

\clearpage
\footnotesize
\bibliography{mybibfile}

\end{document}